\documentclass[final]{siamltex}

\usepackage{pb-diagram}
\usepackage{amsfonts}
\usepackage{latexsym}
\usepackage{amssymb}
\usepackage{amsmath}
\usepackage[dvips]{graphicx}
\usepackage{algorithmic}
\usepackage[ruled]{algorithm}
\usepackage{harvard}

\newcommand{\pf}{\emph{Proof: }}
\newcommand{\ud}{\mathrm{d}}

\newcommand{\call}{\mathcal{L}}

\newcommand{\calu}{\mathcal{U}}

\newcommand{\real}{\mathbb{R}}
\newcommand{\SSS}{\mathbb{S}}

\newcommand{\Yb}{\bar{Y}}

\newcommand{\half}{\frac{1}{2}}

\newcommand{\tY}{\tilde{Y}}
\newcommand{\tX}{\tilde{X}}
\newcommand{\spn}{\mathrm{span}}

\newcommand{\tr}{\mathrm{tr}}
\newcommand{\Yd}{\dot{Y}}
\newcommand{\Ydd}{\ddot{Y}}

\newcommand{\psid}{\dot{\psi}}

\newcommand{\T}{T_{n,d}}

\newcommand{\R}{M_{n,d}^*}

\newcommand{\symn}{\SSS_{n}}
\newcommand{\Sym}{\mathbb{S}_{n}}
\newcommand{\Cnd}{C_{n,d}}

\newcommand{\by}{\mathrm{ \times }}
\newcommand{\Sdmone}{S^{d-1}}
\newcommand{\Tnd}{T_{n,d}}
\newcommand{\Od}{O_{d}}
\newcommand{\Mnd}{M_{n,d}}
\newcommand{\id}{\mathrm{id}}

\newcommand{\Hess}{\mathrm{ Hess}}

\newcommand{\Sp}{S^{d-1}}

\newcommand{\cholnd}{\mathrm{Chol}_{n,d}}
\newcommand{\calH}{{\mathcal{H}}}

\newcommand{\tZ}{\tilde{Z}}
\newcommand{\Pu}{T_{n,d}^*}

\newcommand{\dxi}{\frac{\partial}{\partial x^i}}

\newcommand{\dYij}{\frac{\partial F}{\partial Y_{ij} }}

\newcommand{\Grad}{\mathrm{grad}}

\title{Efficient Rank Reduction of Correlation Matrices\thanks{Date: 12 January 2005. We are grateful for the comments of Erik Balder, Pieter Eendebak, Andrea Giaccobe, Antoon Pelsser, Jordan Rizov and
seminar participants at ECMI Conference 2004 (Eindhoven, The Netherlands), Leiden University, Quantitative
Methods in Finance Conference 2003 (Sydney, Australia), Utrecht University, and Winter School on Mathematical
Finance 2003 (Lunteren, The Netherlands). The second author is grateful for the financial support of the Erasmus
Center of Financial Research.}}
\author{Igor Grubi\v{s}i\'{c}\thanks{Mathematical Institute, University of Utrecht, P.O. Box 80010,
3508 TA Utrecht, The Netherlands (email: grubisic@math.uu.nl, tel: +31 30 2531529)} \and Raoul Pietersz\thanks{Erasmus Research Institute of
Management, Erasmus University Rotterdam, P.O. Box 1738, 3000 DR Rotterdam, The Netherlands (email: pietersz@few.eur.nl, tel: +31 10 4088932)
and Product Development Group (HQ7011), ABN AMRO Bank, P.O. Box 283, 1000 EA Amsterdam, The Netherlands}}

\begin{document}

\maketitle

\begin{abstract}
Geometric optimisation algorithms are developed that efficiently find the nearest low-rank correlation matrix. We show, in numerical tests, that
our methods compare favourably to the existing methods in the literature. The connection with the Lagrange multiplier method is established,
along with an identification of whether a local minimum is a global minimum. An additional benefit of the geometric approach is that any
weighted norm can be applied. The problem of finding the nearest low-rank correlation matrix occurs as part of the calibration of multi-factor
interest rate market models to correlation.
\end{abstract}


\pagestyle{myheadings} \markboth{I. Grubi\v{s}i\'{c} \& R. Pietersz}{Efficient Rank Reduction of Correlation Matrices}
\section{Introduction}

The problem of finding the nearest low-rank correlation matrix occurs in  areas such as finance, chemistry,
physics and image processing. The mathematical formulation of this problem is as follows. Let $\Sym$ denote the
set of real symmetric $n\times n$ matrices and let $C$ be a symmetric $n\by n$ matrix with unit diagonal. For
$X\in\Sym$ we denote by $X \succeq 0$ that $X$ is positive semidefinite. Let the desired rank
$d\in\{1,\dots,n\}$ be given. The problem is then given by
\begin{equation}
\begin{tabular}{rl}
 \qquad \textbf{Find} &   $X \in \Sym$  \\
 \qquad \textbf{to minimize}  &   $ \half \|C-X\|^2 $ \\
  \textbf{subject to}   &   $\rank(X)\leq d$; \ $X_{ii}=1,$  $i=1,\dots,n;$ \ $X \succeq 0$.
\end{tabular}\label{eq:problem}
\end{equation}
Here $\|\cdot\|$ denotes a semi-norm on $\Sym$. The most important instance is
\begin{equation} \label{eq:our semi-norm}
\half \|C-X\|^2 =\half \sum_{i<j} W_{ij}(C_{ij}-X_{ij})^2,
\end{equation}
where $W$ is a weights matrix consisting of non-negative elements. In words: Find the low-rank correlation matrix $X$ nearest to the given $n
\by n$ matrix $C$. The choice of the semi-norm will reflect what is meant by nearness of the two matrices. The semi-norm in (\ref{eq:our
semi-norm}) is well known in the literature, and it is called the \emph{Hadamard semi-norm}, see \citeasnoun{hoj90}. Note that the constraint
set is non-convex for $d<n$, which makes it not straightforward to solve Problem (\ref{eq:problem}) with standard convex optimization methods.

For concreteness, consider the following example. Suppose $C$ is
\[
\left(
\begin{array}{ccc}
    \phantom{-}1.0000 &    -0.1980 &    -0.3827 \\
   -0.1980 &     \phantom{-}1.0000 &    -0.2416 \\
   -0.3827 &    -0.2416 &     \phantom{-}1.0000 \\
\end{array}
\right),
\]
and $W$ is the full matrix, $W_{ij}=1$. With the algorithm developed in this paper, we solve (\ref{eq:problem})
with $C$ as above and $d=2$. The algorithm takes as initial input a matrix $X^{(0)}$ of rank 2 or less, for
example,
\[
X^{(0)}=\left(
\begin{array}{ccc}
    1.0000 &     0.9782 &     0.8982 \\
    0.9782 &     1.0000 &     0.9699 \\
    0.8982 &     0.9699 &     1.0000 \\
\end{array}
\right),
\]
and then produces a sequence of points on the constraint set that converges to the point
\[
X^*=
\left(
\begin{array}{ccc}
   \phantom{-}1.0000 &    -0.4068 &    -0.6277 \\
   -0.4068 &     \phantom{-}1.0000 &    -0.4559 \\
   -0.6277 &    -0.4559 &     \phantom{-}1.0000 \\
\end{array}
\right)
\]
that solves (\ref{eq:problem}). The constraint set and the points generated by the algorithm have been represented in Figure\begin{figure}
\begin{center}
  \includegraphics[width=4in]{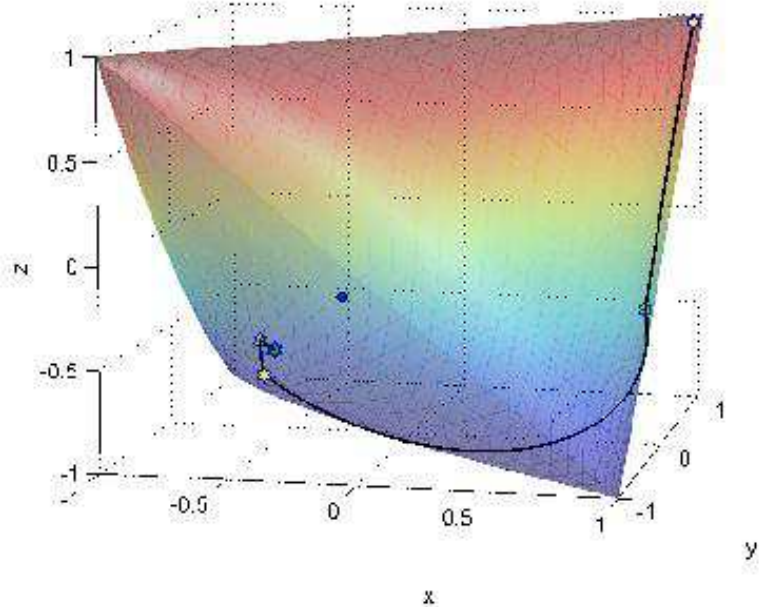}\\
  \caption{The shell represents the set of $3\times 3$ correlation matrices of rank 2 or less. The details of this representation are given in
  Section \ref{sec:case d=2, n=3}.}
   \label{fig:voorbeeld}
  \end{center}
\end{figure}
\ref{fig:voorbeeld}. The details of this representation are given in Section \ref{sec:case d=2, n=3}. The blue
point in the center and the green point represent, respectively, the target matrix $C$ and the solution point
$X^*$. As the figure suggests, the algorithm has fast convergence and the constraint set is a curved space.

This novel technique we propose, is based on geometric optimisation that can locally minimize the objective function in (\ref{eq:problem}) and
which incorporates the Hadamard semi-norm. In fact, our method can be applied to any sufficiently smooth objective function. Not all other
methods available in the literature that aim to solve (\ref{eq:problem}) can handle an arbitrary objective function, see the literature review
in Section \ref{sec:literature review}. We formulate the problem in terms of Riemannian geometry. This approach allows us to use numerical
methods on manifolds that are numerically stable and efficient, in particular the Riemannian-Newton method is applied. We show, for the
numerical tests we performed, that the numerical efficiency of geometric optimisation compares favourably to the other algorithms available in
the literature. The only drawback of the practical use of geometric optimisation is that the implementation is rather involved. To overcome this
drawback, we have made available a MATLAB implementation `LRCM min' (low-rank correlation matrices minimization) at
\texttt{www.few.eur.nl/few/people/pietersz}.

We develop a technique to instantly check whether an obtained local minimum is a global minimum, by adaptation of Lagrange multiplier results of
\citeasnoun{zhw03}. The novelty consists of an expression for the Lagrange multipliers given the matrix $X$, whereas until now only the reverse
direction (an expression for the matrix $X$ given the Lagrange multipliers) was known. The fact that one may instantly identify whether a local
minimum is a global minimum is very rare for non-convex optimisation problems, and that makes Problem (\ref{eq:problem}), which is non-convex
for $d<n$, all the more interesting.

Problem (\ref{eq:problem}) is important in finance, as it occurs as part of the calibration of the multi-factor
LIBOR\footnote{London inter-bank offer rate.} market model of \citeasnoun{bgm97}, \citeasnoun{mss97},
\citeasnoun{jam97} and \citeasnoun{mur97}. This model is an interest rate derivatives pricing model and it is
used in some financial institutions for valuation and risk management of their interest rate derivatives
portfolio. The number of stochastic factors needed for the model to fit to the given correlation matrix is equal
to the rank of the correlation matrix. This rank can be as high as the number of forward LIBORs in the model,
i.e., as high as the dimension of the matrix. The number of LIBORs in the model can grow large in practical
applications, for example, a model with over 80 LIBORs is not uncommon. This implies that the number of factors
needed to fit the model to the given correlation matrix can be high, too. There is much empirical evidence that
the term structure of interest rates is driven by multiple factors (three, four, or even more), see the review
article of \citeasnoun{das03}. Though the number of factors driving the term structure may be four or more, the
empirical work shows that it is certainly not as high as, say, 80. This is one reason for using a model with a
low number of factors. Another reason is the enhanced efficiency when estimating the model-price of an interest
rate derivative through Monte Carlo simulation. First, a lower factor model simply requires drawing less random
numbers than a higher factor model. Second, the complexity of calculating LIBOR rates over a single time step in
a simulation implementation is of order $n\times d$, with $n$ the number of LIBORs and $d$ the number of
factors, see \citeasnoun{jos03}.

The importance of Problem (\ref{eq:problem}) in finance has been recognized by many researchers. In fact, the
literature review of Section \ref{sec:literature review} refers to seventeen articles or books addressing the
problem.

Due to its generality our method finds locally optimal points for a variety of other objective functions subject
to the same constraints. One of the most famous problems comes from physics and is called \emph{Thomson's
problem}. The Thomson problem is concerned with minimizing the potential energy of $n$ charged particles on the
sphere in $\real^3$ ($d = 3$). Geometric optimisation techniques have previously been applied to the Thomson
problem by \citeasnoun{des98}, but these authors have only considered conjugate gradient techniques on a
`bigger' manifold, in which the freedom of rotation has not been factored out. In comparison, we stress here
that our approach considers a lower dimensional manifold, which allows for Newton's algorithm (the latter not
developed in \citeasnoun{des98}). An implementation of geometric optimisation applied to the Thomson problem has
also been included in the `LRCM min' package.

Finally, for a literature review of interest rate models, the reader is referred to \citeasnoun{reb04}.

The paper is organized as follows. In Section \ref{sec:literature review}, the literature is reviewed. In
Section \ref{sec:solution methodology geom prog}, the constraints of the problem are formulated in terms of
differential geometry. We parameterize the set of correlation matrices of rank at most $d$  with a manifold
named the \emph{Cholesky manifold}. This is a canonical space for the optimisation of the arbitrary smooth
function subject to the same constraints. In Section \ref{sec:optimisation over manifolds}, the Riemannian
structure of the Cholesky manifold is introduced. Formulas are given for parallel transport, geodesics, gradient
and Hessian. These are needed for the minimization algorithms, which are made explicit. In Section
\ref{sec:convergence analysis}, we discuss the convergence of the algorithms. In Section \ref{sec:case distance
min}, the application of the algorithms to the problem of finding the nearest low-rank correlation matrix is
worked out in detail. In Section \ref{sec:numerical results}, we numerically investigate the algorithms in terms
of efficiency. Finally, in Section 8, we conclude the paper.

\section{Literature review} \label{sec:literature review}

We mention five algorithms available in the literature for rank reduction of correlation matrices, in reverse chronological order.

First, we mention the majorization approach of \citeasnoun[\emph{b}]{pig04risk}. This reference also contains a literature review covering all
algorithms except for the alternating projections algorithm (discussed below). Majorization can handle an entry-weighted objective function and
is guaranteed to converge to a stationary point. The rate of convergence is sub-linear.

Second, we point out the Lagrange multiplier algorithm of \citeasnoun{zhw03} and \citeasnoun{wu03}. The
correlation matrices produced by this algorithm are not guaranteed to converge to a stationary point of
objective function $F$ in (\ref{eq:problem}). This lack of convergence has been pointed out in
\citeasnoun[Section 2]{pig04} and therefore it is not necessary to repeat the reasons here.

Third, we consider the alternating projections algorithm of \citeasnoun{gru02} and \citeasnoun{mow04}. The
discussion below of alternating projections applies only to the problem of rank reduction of correlation
matrices. The method is based on alternating projections onto the set of $n\times n$ matrices with unit diagonal
and onto the set of $n\times n$ matrices of rank $d$ or less. Both these projections can be efficiently
calculated. For projection onto the intersection of two convex sets, \citeasnoun{dyk83} and \citeasnoun{han88}
have shown that convergence to a minimum can be obtained with alternating projections onto the individual convex
sets if a normal vector correction is applied. Their results do not automatically hold for an alternating
projections algorithm with normal correction for Problem (\ref{eq:problem})\footnote{The algorithm with normal
correction for rank reduction has also been studied in \citeasnoun{wei04}.}, since for $d<n$ the set of $n\times
n$ matrices of rank $d$ or less is non-convex. Indeed, \citeasnoun[Section 2]{pig04} and \citeasnoun{mow04}
report that alternating projections with normal correction may fail in solving Problem (\ref{eq:problem}). The
alternating projections algorithm without normal correction stated in \citeasnoun{gru02} and \citeasnoun{mow04}
however always converges to a feasible point, but not necessarily to a stationary point. In fact, in general,
the alternating projections method without normal correction does not converge to a stationary point. The
algorithm thus does not minimize the objective function in (\ref{eq:problem}), it only selects a feasible point
satisfying the constraints of (\ref{eq:problem}).

The fourth important contribution is due to \citeasnoun{hig02}. The algorithm of \citeasnoun{hig02} is the
alternating projection algorithm with normal correction applied to the case $d=n$, i.e., to the problem of
finding the nearest (possibly full-rank) correlation matrix. Note that the space of positive semidefinite
matrices of rank $n$ is indeed convex, therefore the alternating projections with normal correction method is
guaranteed to converge to the global minimum. Since the case $d < n$ is of primary interest in this paper, the
alternating projections method with normal correction will not be considered in the remainder.

Fifth, we mention the parametrization method of \citeasnoun[1999\emph{b}, 1999\emph{c} (Section 10), 2002
(Section 9), 2004\emph{b} (Sections 20.1--20.4)]{reb99}, \citeasnoun{bri02}, \citeasnoun[Section 6.9]{brm01} and
\citeasnoun{rmb02}. The set of correlation matrices of rank $d$ or less $\{YY^T: Y\in\real^{n\times
d},\diag(YY^T)$$=I\}$ is parameterized by trigonometric functions through spherical coordinates
$Y_i=Y_i(\theta_i)$ with $\theta_i\in\real^{d-1}$. As a result, the objective value $F(Y)$ becomes a function
$F(Y(\theta))$ of the angle parameters $\theta$ that live in $\real^{n\times(d-1)}$. Subsequently, ordinary
non-linear optimisation algorithms may be applied to minimize the objective value $F(Y(\theta))$ over the angle
parameters $\theta$. In essence, this approach is the same as geometric optimisation, except for the key
difference of optimising over $\theta$ versus over $Y$. The major benefit of geometric optimisation over the
parametrization method is as follows. Consider, for ease of exposition, the case of equal weights. The
differential $F_Y$, in terms of $Y$, is given simply as $2\psi Y$, with $\psi=YY^T-C$, see (\ref{eq:dF}) below.
Note that $F_Y=2\psi Y$ can thus be efficiently calculated. The differential $F_\theta$, in terms of $\theta$
however, is $2\psi Y$ multiplied by the differential of $Y$ with respect to $\theta$, by the chain rule of
differentiation. The latter differential is less efficient to calculate since it involves numerous sums of
trigonometric functions.

An initial feasible point can be found by a principal component analysis and a suitable re-scaling. This method is explained in Section
\ref{sec:initial feasible point} below, and is due to \citeasnoun{flu88}. It first appeared in a finance setting in \citeasnoun{sid00} and
\citeasnoun{huw00}. The starting point for the five above mentioned algorithms is the re-scaled PCA initial feasible point.

\subsection{Weighted norms} \label{sec:weighted norms}

We mention two reasons for assigning \emph{non-constant} or \emph{non-homogeneous} weights in the objective function of (\ref{eq:our
semi-norm}). First, in our setting $C$ has the interpretation of measured correlation. It can thus be the case that we are more confident of
specific entries of the matrix $C$. Second, the weighted norm of (\ref{eq:our semi-norm}) has important applications in finance, see, for
example, \citeasnoun{hig02}, \citeasnoun{rebb99} and \citeasnoun{pig04}.

The semi-norm in the objective function $F$ can be (i) a Hadamard semi-norm with arbitrary weights per element
of the matrix, as defined in (\ref{eq:our semi-norm}), or (ii) a weighted Frobenius norm $\|\cdot\|_{F,\Omega}$
with $\Omega$ a positive definite matrix. Here $\|X\|_{F,\Omega}^2=\tr(X\Omega X^T\Omega)$. The weighted
Frobenius norm is, from a practical point of view, by far less transparent than the Hadamard or
weights-per-entry semi-norm (\ref{eq:our semi-norm}). The geometric optimisation theory developed in this paper,
and most of the algorithms mentioned in Section \ref{sec:literature review}, can be efficiently applied to both
cases. The Lagrange multipliers and alternating projections methods however can only be efficiently extended to
the case of the weighted Frobenius norm. The reason is that both these methods need to calculate a projection
onto the space of matrices of rank $d$ or less. Such a projection, for the weighted Frobenius norm, can be
efficiently found by an eigenvalue decomposition. For the Hadamard semi-norm, such an efficient solution is not
available, to our knowledge, and as also mentioned in \citeasnoun[page 336]{hig02}.

\section{Solution methodology with geometric optimisation} \label{sec:solution methodology geom prog}

Note that Problem (\ref{eq:problem}) is a special case of the following more general problem:
\begin{equation}
\begin{tabular}{rl}
 \qquad \textbf{Find} &   $X \in \Sym$  \\
 \qquad \textbf{to minimize}  &   $ \tilde{F}(X) $ \\
  \textbf{subject to}   &   $\rank(X)\leq d$; \ $X_{ii}=1,$  $i=1,\dots,n;$ \ $X \succeq 0$.
\end{tabular}\label{eq:problem with general F}
\end{equation}
In this paper methods will be developed to solve Problem (\ref{eq:problem with general F}) for the case when $\tilde{F}$ is twice continuously
differentiable. In the remainder of the paper, we assume $d>1$, since for $d=1$ the constraint set consists of a finite number ($2^{n-1})$ of
points.

\subsection{Basic idea}

The idea for solving Problem (\ref{eq:problem with general F}) is to parameterize the constraint set by a
manifold, and subsequently utilize the recently developed algorithms for optimisation over manifolds, such as
Newton's algorithm and conjugate gradient algorithms. Such \emph{geometric optimisation} has been developed by
\citeasnoun{smi93}.

In Section \ref{sec:topological structure}, the constraint set is equipped with a topology, and we make an
identification with a certain quotient space. In Section \ref{sec:differentiable structure}, it will be shown
that the constraint set as such is not a manifold; however a dense subset is shown to be a manifold, namely the
set of matrices of rank exactly $d$. Subsequently, in Section \ref{sec:cholesky manifold}, we will define a
larger manifold (named \emph{Cholesky manifold}), of the same dimension as the rank-$d$ manifold, that maps
surjectively to the constraint set. We may apply geometric optimisation on the Cholesky manifold. The connection
between minima on the Cholesky manifold and on the constraint set will be established.

\subsection{Topological Structure} \label{sec:topological structure}

In this section, the set of $n \by n$ correlation matrices of rank $d$ or less is equipped with the subspace topology from $\Sym$. We
subsequently establish a homeomorphism (i.e., a topological isomorphism) between the latter topological space and the quotient space of $n$
products of the $d-1$ sphere over the group of orthogonal transformations of $\real^d$. Intuitively the correspondence is as follows: We can
associate with an $n\by n$ correlation matrix of rank $d$ a configuration of $n$ points of unit length in $\real^d$ such that the inner product
of points $i$ and $j$ is entry $(i,j)$ of the correlation matrix. Any orthogonal rotation of the configuration does not alter the associated
correlation matrix. This idea is developed more rigorously below.

\begin{definition} \label{def:constraint set}The set of symmetric $n\times n$ correlation matrices of rank at most $d$ is defined by
\[
\Cnd = \big\{ \; X\in\symn \; ; \; \diag(X)=I_n,\;\;\rank(X)\le d,\;\; X\succeq 0\big\}.
\]
Here $I_n$ denotes the $n\times n$ identity matrix and $\diag$ denotes the map $\real^{n\times n}\to\real^{n\times n}$, $\diag(X)_{ij} =
\delta_{ij}X_{ij}$, where $\delta_{ij}$ denotes the Kronecker delta.
\end{definition}

The set $\Cnd$ is a subset of $\symn$. The latter space is equipped with the Frobenius norm $\|\cdot\|_F$, which in turn defines a topology. We
equip $\Cnd$ with the subspace topology.

In the following, the product of $n$ unit spheres $\Sdmone$ is denoted by $\Tnd$. Elements of $\Tnd$ are denoted
as a matrix $Y\in\real^{n\times d}$, with each row vector $Y_i$ of unit length. Denote by $\Od$ the group of
orthogonal transformations of $d$-space. Elements of $\Od$ are denoted by a $d\by d$ orthogonal matrix $Q$.

\begin{definition} \label{def:Snd over Od and the maps}We define the following right $\Od$-action\footnote{It is trivially verified that the map thus defined is indeed an $\Od$ smooth
action: $YI_d = Y$ and $Y(Q_1 Q_2)^{-1}=(YQ_2^{-1}){Q_1}^{-1}$. Standard matrix multiplication is smooth.} on $\Tnd$:
\begin{equation} \label{eq:the action}
\Tnd \times \Od \to \Tnd,\;\;(Y, Q)\mapsto YQ.
\end{equation}
An equivalence class $\{YQ:Q \in \Od \}$ associated with $Y\in\Tnd$ is denoted by $[Y]$ and it is called the
orbit of $Y$. The quotient space $\Tnd/\Od$ is denoted by $\Mnd$. The canonical projection
$\Tnd\to\Tnd/\Od=\Mnd$ is denoted by $\pi$. Define the map\footnote{Although rather obvious, it will be shown in
Theorem \ref{thm:the homeomorphism} that this map is well defined.} $\Psi$ as
\[
\Mnd \stackrel{\Psi}{\longrightarrow} \Cnd,\;\;\Psi\big( \; [Y] \; \big)=YY^T.
\]
Consider a map $\Phi$ in the inverse direction of $\Psi$,
\[
\Cnd \stackrel{\Phi}{\longrightarrow} \Mnd,
\]
defined as follows: For $X\in \Cnd$ take $Y\in\Tnd$ such that $YY^T=X$. Such $Y$ can always be found as will be
shown in Theorem \ref{thm:the homeomorphism} below. Then set $\Phi(X)=[Y]$. It will be shown in Theorem
\ref{thm:the homeomorphism} that this map is well defined. Finally, define the map $s:\Tnd\to\Cnd$, $s(Y)=YY^T$.
\end{definition}

The following theorem relates the spaces $\Cnd$ and $\Mnd$; the proof has been deferred to Appendix \ref{sec:app proof thm the homeomorphism}.
\begin{theorem} \label{thm:the homeomorphism}
Consider the following diagram
\begin{equation} \label{eq:the diagram}
\begin{diagram}
\node{\Tnd} \arrow{e,t}{s} \arrow{s,l}{\pi} \node{\Cnd}  \\
\node{\Mnd=\Tnd/\Od} \arrow{ne,t}{\Phi} \node{}
\end{diagram}\hspace{-128pt}
\begin{diagram}
\node{\phantom{\Tnd}} \arrow{e,-,!} \arrow{s,-,!} \node{\phantom{\Cnd}} \arrow{sw,b}{\Psi} \\
\node{\phantom{\Mnd=\Tnd/\Od}} \node{}
\end{diagram}
\end{equation}
with the objects and maps as in Definitions \ref{def:constraint set} and \ref{def:Snd over Od and the maps}. We have the following:
\begin{enumerate}
\item[(i)] The maps $\Psi$ and $\Phi$ are well defined. \item[(ii)] The diagram is commutative, i.e.,
$\Psi\circ\pi=s$ and $\Phi\circ s=\pi$. \item[(iii)] The map $\Psi$ is a homeomorphism with inverse $\Phi$.
\end{enumerate}
\end{theorem}

\subsection{A dense part of $\Mnd$ equipped with a differentiable structure} \label{sec:differentiable structure}

For an exposition on differentiable manifolds, the reader is referred to \citeasnoun{doC92}. It turns out that
$\Mnd$ is not a manifold, but a so-called \emph{stratified space}, see, e.g., \citeasnoun{duk00}. However there
is a subspace of $\Mnd$ that is a manifold, which is the manifold of equivalence classes of matrices of
\emph{exactly }rank $d$. The proof of the following proposition has been deferred to Appendix \ref{sec:app proof
free part manifold}.
\begin{proposition} \label{prop:free part is manifold} Let $\Tnd^*\subset\Tnd$ be the subspace defined by
\[
\Tnd^* = \big\{ \; Y\in\Tnd \; : \; \rank(Y)=d \; \big\}.
\]
Then we have the following:
\begin{enumerate}
\item $\Tnd^*$ is a sub-manifold of $\Tnd$. \item Denote by $\Mnd^*$ the quotient space $\Tnd^*/\Od$. Then $\Mnd^*$ is a manifold of dimension
$n(d - 1)- d(d - 1)/2$.
\end{enumerate}
\end{proposition}
As shown in Proposition \ref{prop:free part is manifold}, a subset $\Mnd^*$ of $\Mnd$ is a manifold. In the
following, we will study charts given by sections of the manifold $\Mnd^*$ that will ultimately lead to the
final manifold over which will be optimized.

A \emph{section} on $\R$ is a map $\sigma:\calu\to \Pu$, with $\calu$ open in $\Mnd^*$, such that $\pi \circ
\sigma=\id_{\R}$. Such a map singles out a unique matrix in each equivalence class. In our case we can
explicitly give such a map $\sigma$. Let $[Y]$ in $\R$, and let $I$ denote a subset of $\{1,\dots,n\}$ with
exactly $d$ elements, such that $\dim(\spn(\{Y_i  :  i \in I\}))=d$, for $Y\in[Y]$. Note that $I$ is well
defined since any two $Y^{(1)},Y^{(2)}\in[Y]$ are coupled by an orthogonal transformation, see the proof of
Theorem \ref{thm:the homeomorphism}, and orthogonal transformations preserve independence. The collection of all
such $I$ is denoted by $\mathcal{I}_Y$. It is readily verified that $\mathcal{I}_Y$ is not empty. Let $\prec$
denote the lexicographical ordering, then $(\mathcal{I}_Y,\prec)$ is a well-ordered set. Thus we can choose the
smallest element, denoted by $J(Y)=(j_1,\dots,j_d)$. Define $\tY\in\real^{d\times d}$ by taking the rows of $Y$
from $J_Y$, thus $\tY_i=Y_{j_i}$. Define $\tX=\tY\tY^T$. Since $\tX$ is positive definite, Cholesky
decomposition can be applied to $\tX$, see for example \citeasnoun[Theorem 4.2.5]{gvl96}, to obtain a unique
lower-triangular matrix $\Yb$ such that $\Yb\Yb^T=\tX$ and $\Yb_{ii}>0$. By Theorem \ref{thm:the homeomorphism},
there exists a unique orthogonal matrix $Q\in O_d$ such that $\Yb=\tY Q$. Define $Y^*=YQ$. Note that $Y^*$ is
lower-triangular, since for $i\notin J_Y$, let $p$ be the largest integer such that $i>j_p$, then $Y_i^*$ is
dependent on $Y_1^*,\dots,Y_{j_p}^*$, as $J_Y$ is the smallest element from $\mathcal{I}_Y$, which implies a
lower-triangular form for $Y^*$. Then define $\mathcal{U}_Y=\{[Z] \ : \ J(Y)\in\mathcal{I}_Z \}\subset \Mnd^*$.
It is obvious that $\mathcal{U}_Y$ and $\pi^{-1}(\mathcal{U}_Y)$ are open in the corresponding topologies. Then
\begin{equation}\label{eq:section}
\sigma_{Y}: \mathcal{U}_Y \to \pi^{-1}(\mathcal{U}_Y), \;\; [Z] \mapsto Z^*,
\end{equation}
is a section of $\mathcal{U}_Y$ at $Y$. The following proposition shows that the sections are the charts of the manifold $\Mnd^*$. The proof has
been deferred to Appendix \ref{sec:app proof diffstructureMnd}.

\begin{proposition}\label{prop:diffstructureMnd}
The differentiable structure on $\R$ is the one which makes $\sigma_Y:\mathcal{U}_Y \to \sigma{(\mathcal{U}_Y)}$
into a diffeomorphism.
\end{proposition}


\subsection{The Cholesky manifold} \label{sec:cholesky manifold}

In this section, we will show that, for the purpose of optimisation, it is sufficient to perform the optimisation on a compact manifold that
contains one of the sections. For simplicity we choose the section $\sigma_Y$ where $J(Y)=\{1, \dots, d\}$. The image $\sigma_Y(\mathcal{U}_Y)$
is a smooth sub-manifold of $\Tnd$ with the following representation in $\real^{n \by d}$
\[
\Big\{ \; \left( \begin{array}{ccc} Y_1^T & \dots & Y_n^T \end{array} \right)^T \; : \ Y_1=(1,0,\dots,0) ; \
Y_i\in S^{i-1}_+,\ i=2,\dots,d; \; Y_i\in\Sp, \ i=d+1,\dots,n \; \Big\},
\]
with $S^{i-1}_+$ embedded in $\real^d$ by the first $i$ coordinates such that coordinate $i$ is bigger than $0$
and with the remaining coordinates set to zero. Also, $S^{d-1}$ is similarly embedded in $\real^d$. We can
consider the map $s:\Tnd\to\Cnd$ restricted to $\sigma_Y(\mathcal{U}_Y)$, which is differentiable since
$\sigma_Y(\mathcal{U}_Y)$ is a sub-manifold of $\Tnd$. The map $s|_{\sigma_Y(\mathcal{U}_Y)}$ is a
homeomorphism, in virtue of Theorem \ref{thm:the homeomorphism}.

For the purpose of optimisation, we need a compact manifold which is surjective with $C_{n,d}$. Define the following sub-manifold of $\Tnd$ of
dimension $n(d - 1)- d(d - 1)/2$,
\[
\cholnd = \Big\{ \; Y\in\real^{n\times d} \; : \ Y_1=(1,0,\dots,0) ; \ Y_i\in S^{i-1},\ i=2,\dots,d; \;
Y_i\in\Sp, \ i=d+1,\dots,n \; \Big\},
\]
which we call the Cholesky manifold. The Cholesky parametrization has been considered before by
\citeasnoun{rmb02}, but these authors do not consider non-Euclidean geometric optimisation. The map
$s|_{\cholnd}$ is surjective, in virtue of the following theorem, the proof of which has been relegated to
Appendix \ref{sec:app proof surjective}.

\begin{theorem} \label{thm:surjective}
If $X\in\Cnd$, then there exists a $Y\in\cholnd$ such that $YY^T=X$.
\end{theorem}

A function $\tilde{F}$ on $\Cnd$ can be considered on $\cholnd$, too, via the composition
\[
\cholnd \stackrel{s}{\to} \Cnd \stackrel{\tilde{F}}{\to} \real, \;\; Y\mapsto YY^T \mapsto \tilde{F}(YY^T).
\]
From here on, we will write $F(Y):=\tilde{F}(YY^T)$ viewed as a function on $\cholnd$.

For a \emph{global} minimum  $F(Y)$ on $\cholnd$, we have that $YY^T$ attains a global minimum of $\tilde{F}$ on $\Cnd$, since the map
$s:\cholnd\to\Cnd$ is surjective. For a \emph{local} minimum,  we have the following theorem. The proof has been deferred to Appendix
\ref{sec:app proof thm loc min on chol is loc min on corr}.
\begin{theorem} \label{thm:loc min on chol is loc min on corr}
The point $Y$ attains a local minimum of $F$ on $\cholnd$ if and only if $YY^T$ attains a local minimum of $\tilde{F}$ on $\Cnd$.
\end{theorem}

These considerations on global and local minima on $\cholnd$ show that, to optimize $\tilde{F}$ over $\Cnd$, we might as well optimize $F$ over
the manifold $\cholnd$. For the optimisation of $\tilde{F}$ over $\Cnd$, there is no straightforward way to use numerical methods such as Newton
and conjugate gradient, since they require a notion of differentiability, but for optimisation of $F$ on $\cholnd$, we can use such numerical
methods.


\subsection{Choice of representation}

In principle, we could elect another manifold $\tilde{M}$ and a surjective open map $\tilde{M}\to \Cnd$. We
insist however on explicit knowledge of the geodesics and parallel transport, for this is essential to obtaining
an efficient algorithm. We found that if we choose the Cholesky manifold then convenient expressions for
geodesics, etc., are obtained. Moreover, the Cholesky manifold has the minimal dimension, i.e.,
$\dim(\cholnd)=\dim(\Mnd^*)$.

In the next section, the geometric optimisation tools are developed for the Cholesky manifold.

\section{Optimisation over the Cholesky manifold} \label{sec:optimisation over manifolds}

For the development of minimization algorithms on a manifold, certain objects of the manifold need to calculated explicitly, such as geodesics,
parallel transport, etc. In this section, these objects are introduced and made explicit for $\cholnd$.

From a theoretical point of view, it does not matter which coordinates we choose to derive the geometrical properties of a manifold. For the
numerical computations however this choice is essential because the simplicity of formulas for the geodesics and parallel transport depends on
the chosen coordinates. We found that simple expressions are obtained when $\cholnd$ is viewed as a sub-manifold of $\Tnd$, which, in turn, is
viewed as a subset of the \emph{ambient space} $\real^{n \by d}$. This representation reveals that, to calculate geodesics and parallel
transport on $\cholnd$, it is sufficient to calculate these on a single sphere.

The tangent space of the manifold $\cholnd$ at a point $Y\in \cholnd$ is denoted by $T_Y \cholnd$. A tangent vector at a point $Y$ is an element
of $T_Y \cholnd$ and is denoted by $\Delta$.

\subsection{Riemannian structure} \label{sec:riemannian structure}

We start with a review of basic concepts of Riemannian geometry. Our exposition follows \citeasnoun{doC92}. Let $M$ be an $m$-dimensional
differentiable manifold. A \emph{Riemannian structure} on $M$ is a smooth map $Y\mapsto\langle\cdot,\cdot\rangle_Y$, which for every $Y \in M$
assigns an inner product $\langle\cdot,\cdot\rangle_Y$ on $T_Y M$, the tangent space at point $Y$. \emph{A Riemannian manifold} is a
differentiable manifold with a Riemannian structure.

Let $F$ be a smooth function on a Riemannian manifold $M$. Denote the \emph{differential} of $F$ at a point $Y$
by $F_Y$. Then $F_Y$ is a linear functional on $T_Y M$. In particular, let $\gamma(t)$,
$t\in(-\varepsilon,\varepsilon)$, be a smooth curve on $M$ such that $\gamma(0)=Y$ and $\dot{\gamma}(0)$
expressed in a coordinate chart $(\calu,x_1,\dots,x_m)$ is equal to $\Delta$, then $F_Y(\Delta)$ can be
expressed in this coordinate chart by
\begin{equation}\label{eq:differential}
F_Y(\Delta)= \sum_{i=1}^m \dxi (F\circ x_i^{-1}) (\gamma)\Big|_{t=0}
\end{equation}
The linear space of linear functionals on $T_Y M$ (the dual space) is denoted by $(T_Y M)^*$. A \emph{vector field} is a map on $M$ that selects
a tangent $\Delta\in T_Y M$ at each point $Y\in M$. The Riemannian structure induces an isomorphism between $T_YM$ and $(T_YM)^{*}$, which
guarantees the existence of a unique vector field on $M$, denoted by $\Grad F$, such that
\begin{equation}\label{eq:grad}
F_{Y}(\Delta)=\langle\Grad F, X \rangle_Y \ \text{ for all } X \in T_Y M.
\end{equation}
This vector field is called the \emph{gradient} of $F$. Also, for Newton and conjugate gradient methods, we have to use second order
derivatives. In particular, we need to be able to differentiate vector fields. To do this on a general manifold, we need to equip the manifold
with additional structure, namely the \emph{connection}. A connection on a manifold $M$ is a rule $\nabla_\cdot\cdot$ which assigns to each two
vector fields $X_1, X_2 $ on $M$ a vector field $\nabla_{X_1} X_2$ on $M$, satisfying the following two conditions:
\begin{equation}\label{eq:affien conn}
\nabla_{FX_1+GX_2}X_3=F\nabla_{X_1}X_3+G \nabla_{X_2}X_3, \;\; \nabla_{X_1}(F X_2)=F\nabla_{X_1}(X_2)+ (X_1 F)X_2,
\end{equation}
for $F,G$ smooth functions on $M$ and $X_1,X_2, X_3 $ vector fields on $M$.

Let $\gamma(t)$ be a smooth curve on $M$ with tangent vector $X_1(t)=\dot{\gamma}(t)$. A given family $X_2(t)$ of tangent vectors at the points
$\gamma(t)$ is said to be \emph{parallel transported} along $\gamma$ if
\begin{equation}\label{eq:parralertransp}
\nabla_{X_1}X_2=0 \ \text{ on } \gamma(t),
\end{equation}
where $X_1, X_2$ are vector fields that coincide with $X_1(t)$ and $X_2(t)$, respectively, on $\gamma(t)$. If the tangent vector $X_1(t)$ itself
is parallel transported along $\gamma(t)$ then the curve $\gamma(t)$ is called a \emph{geodesic}. In particular, if $(\calu,x_1,\dots,x_m)$ is a
coordinate chart on $M$ and $\{X_1, \dots, X_m \}$ the corresponding vector fields then the affine connection $\nabla$ on $\calu$ can be
expressed by
\begin{equation}\label{eq:crisot}
\nabla_{X_i}X_j=\sum_{k=1}^{m} \Gamma^{k}_{i,j}X_k.
\end{equation}
The functions $\Gamma^{k}_{i,j}$ are  smooth functions, called the \emph{Christoffel symbols} for the connection. In components, the geodesic
equation becomes
\begin{equation}\label{eq:geoddef}
\ddot{x}_{k} + \sum_{i,j=1}^m\Gamma^{k}_{i,j}\dot{x}_{i}\dot{x}_{j}=0,
\end{equation}
where $x_k$ are the coordinates of $\gamma(t)$. On a Riemannian manifold there is a unique torsion free connection compatible with the metric,
called the \emph{Levi-Civita} connection. This means that Christoffel symbols can be expressed as  functions of a metric on $M$. Note also that
(\ref{eq:geoddef}) implies that, once we have determined the equation for the geodesic we can simply read off Christoffel symbols. With respect
to an induced metric the geodesic is the curve of shortest length between two points on a manifold. For a manifold embedded in Euclidean space
an equivalent characterization of a geodesic is that the acceleration vector at each point along a geodesic is normal to the manifold so long as
the curve is traced with uniform speed.

We start by defining Riemannian structures for $\Tnd$ and for the Cholesky manifold $\cholnd$. We use the Levi-Civita connection, associated to
the metric defined as follows on the tangent spaces. Both tangent spaces are identified with suitable subspaces of the ambient space
$\real^{n\times d}$, and subsequently the inner product for two tangents $\Delta_1$, $\Delta_2$ is defined as
\begin{equation} \label{eq:riemannian structure}
\langle \Delta_1,\Delta_2 \rangle = \tr \Delta_1 \Delta_2^T,
\end{equation}
which is the Frobenius inner product for $n\times d$ matrices. Note that, in our special case, the inner product $\langle\cdot,\cdot\rangle_Y$
is independent of the point $Y$; therefore we suppress the dependency on $Y$.

\subsection{Normal and tangent spaces}

An equation determining tangents to $\T$ at a point $Y$ can be obtained by differentiating $\diag(YY^T)=I_n$\,
yielding $\diag(Y\Delta^T +\Delta Y^T)=0$, i.e., $\diag(\Delta Y^T)=0$. The dimension of the tangent space is
$n(d-1)$. The \emph{normal space} at the point $Y$ is defined to be the orthogonal complement of the tangent
space at the point $Y$, i.e., it consists of the matrices $N$, for which $\tr\Delta N^T=0$ for all $\Delta$ in
the tangent space.  It follows that the normal space is $n$ dimensional.  It is straightforward to verify that
if $N=DY$, where $D$ is $n\by n$ diagonal, then $N$ is in the normal space. Since the dimension of the space of
such matrices is $n$, we see that the normal space  $N_Y \T$ at $Y\in \T$ is given by
\[
N_Y \T= \big\{ \; D Y \; ; \; D \in\real^{n\times n} \; \textrm{ diagonal } \; \big\}.
\]
The projections $\pi_{N_Y\T}$ and $\pi_{T_Y\T}$ onto the normal and tangent spaces of $\T$ are given by
\[
\pi_{N_Y\T}(\Delta) = \diag(\Delta Y^T) Y \;\textrm{ and }\; \pi_{T_Y\T}(\Delta) = \Delta-\diag(\Delta Y^T) Y,
\]
respectively. The projection $\pi_{T_Y\cholnd}$ onto the tangent space of $\cholnd$ is given by
\begin{equation}
\pi_{T_Y\cholnd}(\Delta)= \zeta\big( \ \pi_{T_Y\T}(\Delta)\ \big) \;\textrm{ with }\; \zeta(\Delta)\;\textrm{ defined by }\;
\zeta(\Delta)_{ij}=\left\{
\begin{array}{cl}
0 & \textrm{ for } j>i  \textrm{ or } i=j=1,  \\
\Delta_{ij} & \textrm{ otherwise. }
\end{array}\right.
\end{equation}

\subsection{Geodesics}

It is convenient to work with the coordinates of the ambient space $\real^{n \by d}$. In this coordinate system, geodesics on $\Tnd$ with
respect to the Levi-Civita connection obey the second order differential equation
\begin{equation} \label{eq:Gamma}
\ddot{Y}+\Gamma_Y(\dot{Y},\dot{Y})=0, \ \textrm{ with }
\Gamma_Y(\Delta_1,\Delta_2):=\diag(\Delta_1\Delta_2^T)Y.
\end{equation}

To see this, we begin with the condition that $Y(t)$ remains on $\T$,
\begin{equation}
\label{eq:g0} \diag(YY^T)=I_n.
\end{equation}
Differentiating this equation twice, we obtain,
\begin{equation} \label{eq:g1}
\diag(\Ydd Y^T+2\Yd\Yd^T+Y\Ydd^T)=0.
\end{equation}
In order for $Y(\cdot)$ to be a geodesic, $\Ydd(t)$ must be in the normal space at $Y(t)$, i.e.,
\begin{equation} \label{eq:g2}
\Ydd(t)=D(t)Y(t)
\end{equation}
for some diagonal matrix $D(t)$. To obtain an expression for $D$, substitute (\ref{eq:g2}) into (\ref{eq:g1}), which yields (\ref{eq:Gamma}).

The function $\Gamma_Y$ is the matrix notation of the Christoffel symbols, $\Gamma^k_{ij}$, with respect to $E_{1},\dots,E_{nd}$, the standard
basis vectors of $\real^{n \by d}$. More precisely, $\nabla_{E_{i}}E_{j}=\sum_{k=1}^{nd}\Gamma_{ij}^k E_k$ with $\Gamma_{ij}^k$ defined by
$\langle \Gamma_Y ( X_1 , X_2 ), E_k $ $ \rangle = \sum_{i,j=1}^{nd} \Gamma^k_{ij}(X_1)_i (X_2)_j $.

The geodesic at $Y(0) \in \Tnd$ in the direction $\Delta \in T_{Y(0)}\Tnd$ is given by,
\begin{equation}\label{eq:geodesicstor}
Y_{i}(t)=\cos( \; \| \Delta_i \| t \; \big) \; Y_i(0)+ \frac{1}{\| \Delta_i \|} \sin\big( \; \| \Delta_i \| t \; \big) \; \Delta_{i}.
\end{equation}
for $i=1,\dots,n$, per component on the sphere. By differentiating, we obtain an expression for the evolution of the tangent along the geodesic:
\begin{equation}\label{eq:geodesic tangent}
\Yd_{i}(t)= - \| \Delta_i \|  \sin\big( \; \| \Delta_i \| t \;\big) \; Y_i(0) +  \cos\big( \;\| \Delta_i \| t\; \big) \; \Delta_{i}.
\end{equation}
Since $\cholnd$ is a Riemannian sub-manifold of $\Tnd$ it has the same geodesics.

\subsection{Parallel transport along a geodesic}

We consider this problem per component on the sphere. If $\Delta^{(2)}\in T_{Y^{(1)}}\Tnd$ is parallel transported along a geodesic starting
from $Y^{(1)}$ in the direction of $\Delta^{(1)}\in T_{Y^{(1)}}\Tnd$, then decompose $\Delta^{(2)}$ in terms of $\Delta^{(1)}$,
\[
\Delta_i^{(2)}(t) = \langle \Delta_i^{(1)}(0),\Delta_i^{(2)}(0) \rangle \Delta_i^{(1)}(t) + R_i, \;\; R_i \bot \Delta_i^{(1)}(0).
\]
Then $\Delta_i^{(1)}(t)$ changes according to (\ref{eq:geodesic tangent}) and $R_i$ remains unchanged. Parallel transport from $Y^{(1)}$ to
$Y^{(2)}$ defines an isometry $\tau(Y^{(1)},Y^{(2)}):T_{Y^{(1)}}\Tnd\to T_{Y^{(2)}}\Tnd$. When it is clear in between which two points is
transported, then parallel transport is denoted simply by $\tau$. Since $\cholnd$ is a Riemannian sub-manifold of $\Tnd$ it has the same
equations for parallel transport.

\subsection{The gradient}

Since $\cholnd$ is a sub-manifold of $\real^{n\times d}$ we can use coordinates of $\real^{n \by d}$ to express the differential $F_Y$ of $F$ at
the point $Y$, namely $(F_Y)_{ij}=\dYij $. The gradient $\Grad F$ of a function $F$ on  $\cholnd$ can be determined by ($\ref{eq:grad}$). It
follows that,
\begin{eqnarray}
\Grad F &=& \pi_{T_Y\cholnd}(F_Y) = \zeta\big( \; F_Y - \diag( F_Y Y^T ) Y \; \big). \label{eq:gradient}
\end{eqnarray}

\subsection{Hessian} \label{sec:hessian}

The Hessian $\Hess F$ of a function $F$ is a second covariant derivative of $F$. More precisely, let $\Delta_1, \Delta_2$ be two vector fields,
then
\[
\Hess F (\Delta_1,\Delta_2)= \langle \nabla_{\Delta_1}\grad F,\Delta_2 \rangle
\]
In local coordinates of $\real^{n \by d}$
\begin{equation}\label{eq:hessian} \Hess F(\Delta_1,\Delta_2) =
F_{YY}(\Delta_1,\Delta_2)-\langle F_Y,\Gamma_Y(\Delta_1,\Delta_2)
\rangle,
\end{equation}
where
\[
F_{YY}(\Delta_1,\Delta_2)=\frac{\ud}{\ud t}\frac{\ud}{\ud s}\bigg|_{t=s=0} F(Y(t,s)), \textrm{ with }\frac{\ud}{\ud t}\bigg|_{t=0} Y = \Delta_1,
\;\;\frac{\ud}{\ud s}\bigg|_{s=0} Y = \Delta_2.
\]
Newton's method requires inverting the Hessian at minus the gradient, therefore we need to find the tangent $\Delta$ to $\cholnd$ such that
\begin{equation} \label{eq:Newton requirement}
\Hess F(\Delta,X)=\langle -\Grad F, X\rangle,\textrm{ for all tangents }X\textrm{ to }\cholnd.
\end{equation}
To solve (\ref{eq:Newton requirement}), it is convenient to calculate the unique tangent vector $\calH=\calH(\Delta)$ satisfying
\[
\Hess F(\Delta,X)=\langle \calH, X\rangle,\textrm{ for all tangents }X\textrm{ to }\cholnd,
\]
since then the Newton Equation (\ref{eq:Newton requirement}) becomes $\calH(\Delta)=-\Grad F$. From (\ref{eq:Gamma}) and (\ref{eq:hessian}), we
obtain
\begin{equation}
\calH(\Delta)=\pi_{T_Y\cholnd}(F_{YY}(\Delta))-\diag(F_Y Y^T)\Delta,
\end{equation}
where the notation $F_{YY}(\Delta)$ means the tangent vector satisfying
\[
F_{YY}(\Delta)= \frac{\ud}{\ud t}\bigg|_{t=0}F_Y(Y(t)),\;\; \dot{Y}(0)=\Delta.
\]

\subsection{Algorithms}

We are now in a position to state the conjugate gradient algorithm, given as Algorithm \ref{al:alg2}, and the
Newton algorithm, given as Algorithm \ref{al:alg1}, for optimisation over the Cholesky manifold. These
algorithms are instances of the geometric programs presented in \citeasnoun{smi93}, for the particular case of
the Cholesky manifold.

\section{Discussion of convergence properties} \label{sec:convergence analysis}

\noindent In this section, we discuss convergence properties of the geometric programs: global convergence and
the local rate of convergence.

\subsection{Global convergence}

First, we discuss global convergence for the Riemannian-Newton algorithm. It is well known that the Newton
algorithm, as displayed in Algorithm \ref{al:alg1}, is not globally convergent to a local minimum. Moreover, the
steps in Algorithm \ref{al:alg1} may even not be well defined, because the Hessian mapping could be singular.
The standard way to resolve these issues, is to introduce jointly a steepest descent algorithm. So Algorithm
\ref{al:alg1} is adjusted in the following way. When the new search direction $\Delta^{(k)}$ has been
calculated, then we also consider the steepest descent search direction $\Delta_{\textrm{Steep}}^{(k)}=-\Grad
F(Y^{(k)})$. Subsequently, a line minimization of the objective value is performed in both directions,
$\Delta^{(k)}$ and $\Delta_{\textrm{Steep}}^{(k)}$. We then take as the next point of the algorithm whichever
search direction finds the point with lowest objective value. Such a steepest descent method with line
minimization is well known to have guaranteed convergence to a local minimum.

Second, we discuss global convergence for conjugate gradient algorithms. For the Riemannian case, we have not
seen any global convergence results for conjugate gradient algorithms in the literature. Therefore we focus on
the results obtained for the flat-Euclidean case. \citeasnoun{zou70} and \citeasnoun{alb85} establish global
convergence of the \citeasnoun{flr64} conjugate gradient method with line minimization. \citeasnoun{gin92}
establish alternative line search minimizations that guarantee global convergence of the \citeasnoun{por69}
conjugate gradient method.\begin{algorithm}[tb]
  \caption{Conjugate gradient for minimizing $F(Y)$ on
$\cholnd$ } \label{al:alg2}
 Input: $Y^{(0)},F(\cdot)$.
\begin{algorithmic}[10]
    \REQUIRE  $Y^{(0)} $ such that $Y^{(0)}(Y^{(0)})^T=I_n$.
    \medskip

\STATE{Compute $G^{(0)}=\Grad F(Y^{(0)})=\zeta(F_Y-\diag(F_Y (Y^{(0)})^T) Y^{(0)})$ and set $H^{(0)}=-G^{(0)}$.}

     \FOR{$k=0,1,2,...$}

     \STATE{Minimize
      $F\big(Y^{(k)}(t)\big)$ over~$t$ where
      $Y^{(k)}(t)$ is a geodesic on $\cholnd$ starting from $Y^{(k)}$ in the
      direction of $H^{(k)}$.}

    \STATE{Set $t_k=t_{\min}$ and $Y^{(k+1)}=Y^{(k)}(t_k)$.}

    \STATE{Compute $G^{(k+1)}=\Grad F(Y^{(k+1)})=\zeta\big( \ F_Y-\diag(F_Y (Y^{(k+1)})^T) Y^{(k+1)} \ \big)$.}

    \STATE{Parallel transport tangent vectors $H^{(k)}$ and $G^{(k)}$
      to the point $Y^{(k+1)}$.}

    \STATE{ Compute the new search direction
    \[
        H^{(k+1)}
      =-G^{(k+1)} +\gamma_k\tau H^{(k)}\;\hbox{ where }\;
      \left\{ \begin{array}{ll}
      \gamma_k = {\langle
      G^{(k+1)} -\tau G^{(k)},G^{(k+1)}\rangle\over \langle
      G^{(k)},G^{(k)}\rangle}, & \hbox{\citeasnoun{por69}}, \\[8pt]
      \gamma_k = \frac{||G^{(k+1)}||^2}{||G^{(k)}||^2}, & \hbox{\citeasnoun{flr64}}. \\
      \end{array} \right.
    \]

      }

    \STATE{Reset $H^{(k+1)}=-G^{(k+1)}$ if $k+1\equiv0\mod{n(d-1)-\half d(d-1)}$.}
    \ENDFOR
  \end{algorithmic}
\end{algorithm}\begin{algorithm}[tb]
  \caption{Newton's method for minimizing
$F(Y)$ on $\cholnd$.} \label{al:alg1}
 Input: $Y^{(0)},F(\cdot)$.
\begin{algorithmic}[10]
    \REQUIRE  $Y^{(0)}$ such that $\diag(Y^{(0)}(Y^{(0)})^T)=I_n$.

 \FOR{$k=0,1,2,...$}
 \STATE{Compute $G^{(k)}=\Grad F(Y^{(k)})=\zeta\big( \ F_Y-\diag(F_Y Y^T) Y \ \big)$}.
 \STATE{Compute $\Delta^{(k)}=-\calH^{-1}G^{(k)}$, i.e. $\Delta^{(k)}\in T_Y\cholnd$ and
\[
\zeta\Big( \ F_{YY}(\Delta^{(k)})-\diag\big( \ F_{YY}(\Delta^{(k)}) (Y^{(k)})^T \ \big) Y^{(k)} \ \Big)
-\diag(F_Y (Y ^{(k)})^T)\Delta^{(k)} =-G^{(k)}.
\]
}
 \STATE{Move from $Y^{(k)}$ in direction $\Delta^{(k)}$ to~$Y^{(k)}(1)$ along the geodesic.}
 \STATE{Set $Y^{(k+1)}=Y^{(k)}(1)$.}
 \ENDFOR
  \end{algorithmic}
\end{algorithm}

\subsection{Local rate of convergence}

Local rates of convergence for geometric optimisation algorithms are established in \citeasnoun{smi93},
\citeasnoun{eas99} and \citeasnoun{dpm93}.

In Theorem 3.3 of \citeasnoun{smi93}, the following result is established for the Riemannian-Newton method. If
$\hat{Y}$ is a non-degenerate stationary point, then there exists an open set $\mathcal{U}$ containing
$\hat{Y}$, such that starting from any $Y^{(0)}$ in $\mathcal{U}$, the sequence of points produced by Algorithm
\ref{al:alg1} converges quadratically to $\hat{Y}$.

In Theorem 4.3 of \citeasnoun{smi93}, the following result is stated for the Riemannian \citeasnoun{flr64} and
\citeasnoun{por69} conjugate gradient methods. Suppose $\hat{Y}$ is a non-degenerate stationary point such that
the Hessian at $\hat{Y}$ is positive definite. Suppose $\{Y^{(j)}\}_{j=0}^\infty$ is a sequence of points,
generated by Algorithm \ref{al:alg2}, converging to $\hat{Y}$. Then, for sufficiently large $j$, the sequence
$\{Y^{(j)}\}_{j=0}^\infty$ has $\dim(\cholnd)$-steps quadratic convergence to $\hat{Y}$.

As a numerical illustration, convergence runs have been displayed in Figure\begin{figure}
\begin{center}
  \includegraphics[width=4in]{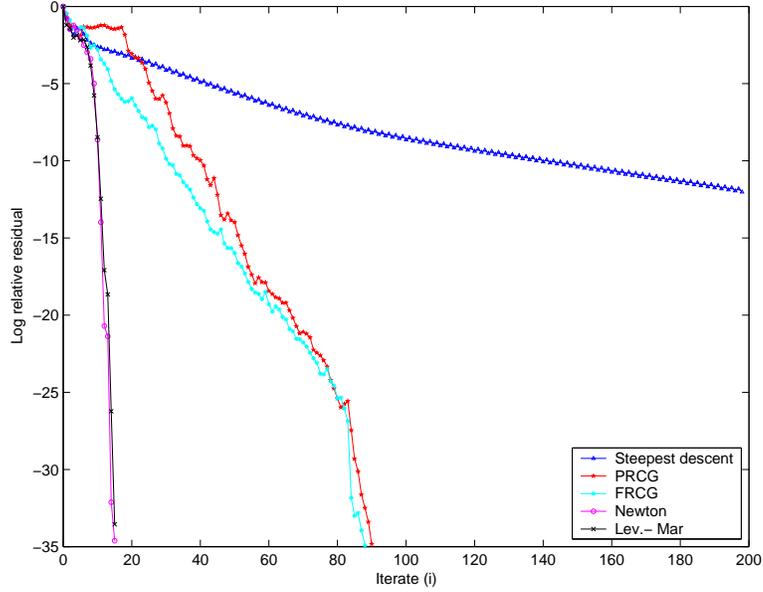}\\
\caption{Convergence runs for various geometric programs: log relative residual $ln$$(\|\Grad F(Y^{(i)})\|$$/\|\Grad F(Y^{(0)})\|)$ versus the
iterate $i$.}
   \label{fig:convergence run}
  \end{center}
\end{figure}
\ref{fig:convergence run}, for reducing a $10\times 10$ correlation matrix to rank 3. The following algorithms
are compared:
\begin{enumerate}
\item[1.] Steepest descent, for which the search direction $H^{(k+1)}$ in Algorithm \ref{al:alg2} is equal to
$-G^{(k+1)}$, i.e., to minus the gradient. The steepest descent method has a linear local rate of convergence,
see \citeasnoun[Theorem 2.3]{smi93}. \item[2.] PRCG, Polak-Ribi\`{e}re conjugate gradient. \item[3.] FRCG,
Fletcher-Reeves conjugate gradient. \item[4.] Newton. \item[5.] Lev.-Mar., the \citeasnoun{lev44} \&
\citeasnoun{mar63} method, which is a Newton-type method.
\end{enumerate}
The code that is used for this test is the package `LRCM min', to be discussed in Section \ref{sec:numerical
results}. This package also contains the correlation matrix used for the convergence run test. Figure
\ref{fig:convergence run} clearly illustrates the convergence properties of the various geometric programs. The
efficiency of the algorithms is studied in Section \ref{sec:numerical results} below.

\section{A special case: Distance minimization} \label{sec:case distance min}

In this section, the primary concern of this paper to minimize the objective function of (\ref{eq:our semi-norm}) is studied. The outline of
this section is as follows. First, some particular choices for $n$ and $d$ are examined. Second, the differential and Hessian of $F$ are
calculated. Third, the connection with Lagrange multipliers is stated; in particular, this will lead to an identification method of whether a
local minimum is a global minimum. Fourth, we discuss the PCA with re-scaling method for obtaining an initial feasible point.

\subsection{The case of $d=n$}

The case that $C$ is a symmetric matrix and the closest positive semidefinite matrix $X$ is to be found allows a
successive projection solution, which was shown by \citeasnoun{hig02}.

\subsection{The case of $d=2$, $n=3$} \label{sec:case d=2, n=3}

A $3\times 3$ symmetric matrix with ones on the diagonals is denoted by
\[
\left(
\begin{array}{ccc}
1 & x & y\\
x & 1 & z\\
y & z & 1
\end{array}
\right).
\]
Its determinant is given by
\[
\det = -\big\{ \; x^2+y^2+z^2 \; \big\} +2xyz+1.
\]
By straightforward calculations it can be shown that $\det=0$ implies that all eigenvalues are nonnegative. The set of 3 by 3 correlation
matrices of rank 2 may thus be represented by the set $\{\det=0\}$. To get an intuitive understanding of the complexity of the problem, the
feasible region has been displayed in Figure \ref{fig:voorbeeld}.

\subsection{Formula for the differential of $F$}

Consider the specific case of the weighted Hadamard semi-norm of (\ref{eq:our semi-norm}). This semi-norm can be represented by a Frobenius norm
by introducing the Hadamard product $\circ$. The Hadamard product denotes entry-by-entry multiplication. Formally, for two matrices $A$ and $B$
of equal dimensions, the Hadamard product $A\circ B$ is defined by $(A\circ B)_{ij}=A_{ij} B_{ij}$. The objective function (\ref{eq:our
semi-norm}) can then be written as
\[
F(Y) = \half \sum_{i<j} W_{ij}(C_{ij}-Y_{i}Y_{j}^T)^2 = \half\|W^{\circ 1/2}\circ \psi\|_F^2 = \half\langle W^{\circ 1/2}\circ\psi, W^{\circ
1/2}\circ\psi \rangle ,
\]
with $\psi := YY^T - C$ and with $(W^{\circ 1/2})_{ij}=\sqrt{W_{ij}}$. Then
\begin{eqnarray*}
\frac{\ud}{\ud t} F(Y(t)) &=& \langle W^{\circ 1/2}\circ\psid, W^{\circ 1/2}\circ\psi \rangle = \langle \psid, W\circ\psi \rangle\\
&=& \langle \Delta Y^T + Y \Delta^T, W\circ\psi \rangle \\
&=& \langle \Delta Y^T, W\circ\psi \rangle + \langle Y \Delta^T, W\circ\psi \rangle \\
&=& \langle \Delta, 2 (W\circ\psi) Y \rangle \; = \; \langle \Delta , F_Y \rangle, \;\; \forall \Delta.
\end{eqnarray*}
Thus from (\ref{eq:differential}) we have
\begin{equation}\label{eq:dF}
F_Y = 2 (W\circ\psi) Y.
\end{equation}
Similarly, we may compute the second derivative
\[
F_{YY}(\Delta) = \frac{\ud }{ \ud t}\Big|_{t=0} F_Y(Y(t)) = 2\Big( (W\circ\psi) \Delta + \big(W\circ(\Delta Y^T+Y \Delta^T )\big) Y \Big),
\]
with $Y(\cdot)$ any curve starting from $Y$ in the direction of $\Delta$.

\subsection{Connection normal with Lagrange multipliers}

The following lemma provides the basis for the connection of the normal vector at $Y$ versus the Lagrange multipliers of the algorithm of
\citeasnoun{zhw03} and \citeasnoun{wu03}. The result is novel since previously only an expression was known for the matrix $Y$ given the
Lagrange multipliers. The result below establishes the reverse direction. This Lagrange result will allow us to identify whether a local minimum
is also a global minimum. That we are able to efficiently determine whether a local minimum is a global minimum, is a very rare phenomenon in
non-convex optimisation, and makes the rank reduction problem (non-convex for $d<n$) all the more interesting.

Note that the Lagrange theory is based on an efficient expression of the low-rank projection by an eigenvalue decomposition. Therefore the
theory below can be extended efficiently only for the Hadamard norm with equal weights and for the weighted Frobenius norm, see also the
discussion in Section \ref{sec:weighted norms}. The proof of the following lemma has been deferred to Appendix \ref{sec:app proof connection
normal lagrange}.
\begin{lemma}
\label{lem:connection normal vs lagrange}Let $Y\in\Tnd$ be such that $\Grad F(Y)=0$. Here, $\Grad F$ is the gradient of $F$ on $\Tnd$, $\Grad
F(Y)=\pi_{T_Y\Tnd}(F_Y)=F_Y - \diag( F_Y Y^T ) Y$, with $F_Y$ in (\ref{eq:dF}). Define
\[
\lambda :=\half \; \diag\big(\; F_YY^T \;\big)
\]
and define $C(\lambda):=C+\lambda$. Then there exist a joint eigenvalue decomposition
\[
C(\lambda) = Q D Q^T, \;\; YY^T = Q D^* Q^T
\]
where $D^*$ can be obtained by selecting at most $d$ nonnegative entries from $D$ (here if an entry is selected it retains the corresponding position in the
matrix).
\end{lemma}

The characterization of the global minimum for Problem (\ref{eq:problem}) was first achieved in \citeasnoun{zhw03} and \citeasnoun{wu03}, which
we repeat here: Denote by $\{X\}_d$ a matrix obtained by eigenvalue decomposition of $X$ together with leaving in only the $d$ largest
eigenvalues (in absolute value). Denote for $\lambda\in\real^n$: $C(\lambda)=C+\diag(\lambda)$. The proof of the following theorem has been
repeated for clarity in Appendix \ref{sec:app proof thm global minimum}.
\begin{theorem}
\label{thm:global minimum}\emph{(Characterization of the global minimum of Problem (\ref{eq:problem}), see \citeasnoun{zhw03} and \citeasnoun{wu03})}
Let $C$ be a symmetric matrix. Let $\lambda^*$ be such that there exists $\{C+\diag(\lambda^*)\}_d \in \Cnd$ with
\begin{equation} \label{eq:assume diag C+l_d = diag C}
\diag\big( \; \{ C+\diag(\lambda^*) \}_d \; \big) = \diag( C ).
\end{equation}
Then $\{C+\diag(\lambda^*)\}_d$ is a global minimizer of Problem (\ref{eq:problem}).
\end{theorem}

This brings us in a position to identify whether a local minimum is a global minimum:
\begin{theorem} \label{thm:when local min is global min}
Let $Y\in\Tnd$ be such that $\Grad F(Y)=0$ on $\Tnd$. Let $\lambda$ and $C(\lambda)$ be defined as in Lemma \ref{lem:connection normal vs
lagrange}. If $YY^T$ has the $d$ largest eigenvalues from $C(\lambda)$ (in absolute value) then $YY^T$ is a global minimizer to the Problem
(\ref{eq:problem}).
\end{theorem}

\pf Apply Lemma \ref{lem:connection normal vs lagrange} and Theorem \ref{thm:global minimum}.\hfill$\Box$

\subsection{Initial feasible point} \label{sec:initial feasible point}

To obtain an initial feasible point $Y\in\Tnd$ we use a method of \citeasnoun{flu88}. We first perform an eigenvalue decomposition
\begin{equation} \label{eq:svd}
C=Q\Lambda Q^T, \;\; \Lambda_{11} \ge \dots \ge \Lambda_{nn}.
\end{equation}
Then we define $Y$ by assigning to each row
\[
Y_i = \frac{Z}{\|Z\|_2}, Z =  (Q_d \Lambda_{d+}^{1/2} )(i,:),\;\; i=1,\dots,n,
\]
where $Q_d$ consists of the first $d$ columns of $Q$ and where $\Lambda_{d+}$ is the principal sub-matrix of $\Lambda$ of degree $d$, filled
only with the non-negative elements from $\Lambda$. The scaling is to ensure that each row of $Y$ is of unit length. If row $i$ is a priori of
zero length, then we choose $Y_i$ to be an arbitrary vector in $\real^d$ of unit length. Finally, to obtain an initial feasible point in
$\cholnd$, we perform a Cholesky decomposition as in the proof of Theorem \ref{thm:surjective}.

Note that the condition of decreasing norm in (\ref{eq:svd}) is thus key to ensure that the initial point is close to the global minimum, see
the result of Theorem \ref{thm:when local min is global min}.

\section{Numerical results} \label{sec:numerical results}

There are many different algorithms available in the literature, as detailed in Section \ref{sec:literature review}. Some of these have an
efficient implementation, i.e., the cost of a single iteration is low. Some algorithms have fast convergence, for example, the Newton method has
quadratic convergence. Algorithms with fast convergence usually require less iterations to attain a predefined convergence criterion. Thus, the
real-world performance of an algorithm is a trade-off between cost-per-iterate and number of iterations required. A priori, it is not clear
which algorithm will perform best. Therefore, in this section, the numerical performance of geometric optimisation is compared to other methods
available in the literature.

\subsection{Acknowledgement} Our implementation of geometric optimisation over low-rank correlation matrices `LRCM min'\footnote{LRCM min can be downloaded from
\texttt{www.few.eur.nl/few/people/pietersz}.} is an adoption of the `SG min' template of \citeasnoun{edl00}
(written in MATLAB) for optimisation over the Stiefel and Grassmann manifolds. This template contains four
distinct well-known non-linear optimisation algorithms adapted for geometric optimisation over Riemannian
manifolds: Newton algorithm; dogleg step or \citeasnoun{lev44} and \citeasnoun{mar63} algorithm;
\citeasnoun{por69} conjugate gradient; and \citeasnoun{flr64} conjugate gradient.

\subsection{Numerical comparison}

The performances of the following seven algorithms, all of these described in Sections \ref{sec:optimisation
over manifolds} and \ref{sec:literature review}, except for item 7 (\texttt{fmincon}), are compared:
\begin{enumerate}
\item Geometric optimisation, Newton (Newton).

\item Geometric optimisation, Fletcher-Reeves conjugate gradient (FRCG).

\item Majorization, e.g., \citeasnoun{pig04} (Major.).

\item Parametrization, e.g., \citeasnoun{reb99jcf} (Param.).

\item Alternating projections without normal vector correction, e.g., \citeasnoun{gru02} (Alt.~Proj.).

\item Lagrange multipliers, e.g., \citeasnoun{zhw03} (Lagrange).

\item \texttt{fmincon}, a MATLAB built-in medium-scale constrained nonlinear program (fmincon).
\end{enumerate}
Note that the first two algorithms in this list have been developed in this paper. The algorithms are tested on a large number of randomly
generated correlation matrices. The benefit of testing on many correlation matrices is, that the overall and generic performance of the
algorithms may be assessed. The correlation matrices are randomly generated as follows. A parametric form for (primarily interest rate)
correlation matrices is posed in \citeasnoun[Equation (8)]{djdp02}. We repeat the parametric form here for completeness.
 \[
 \rho(T_i,T_j) = \exp\Big\{ \; -\gamma_1|T_i-T_j| \; -\frac{\gamma_2|T_i-T_j|}{\max(T_i,T_j)^{\gamma_3}} \; -\gamma_4\big|\sqrt{T_i}-\sqrt{T_j}\big| \;
 \Big\},
 \]
with $\gamma_1, \gamma_2, \gamma_4 > 0$ and with $T_i$ denoting the expiry time of rate $i$. (Our particular choice is $T_i=i$, $i=1,2,\dots$)
This model was then subsequently estimated with USD historical interest rate data. In Table 3 of \citeasnoun{djdp02} the estimated
$\gamma$-parameters are listed, along with their standard error. An excerpt of this table has been displayed in Table \ref{tbl:estimates gamma}.
 \begin{table}  \caption{Excerpt of Table 3 in De Jong et al.~(2004).} 
 \begin{center}
\begin{tabular}{ccccc}
\hline
           &   $\gamma_1$ &   $\gamma_2 $&   $\gamma_3$ &   $\gamma_4$ \\
\hline
  estimate &      0.000 &      0.480 &      1.511 &      0.186 \\
standard error &           -    &      0.099 &      0.289 &      0.127 \\
\hline
\end{tabular}
\end{center}\label{tbl:estimates gamma}
\end{table}
The random correlation matrix that we use is obtained by randomizing the $\gamma$-parameters. We assume the $\gamma$-parameters distributed
normally with mean and standard errors given by Table \ref{tbl:estimates gamma}, with $\gamma_1, \gamma_2, \gamma_4$ capped at zero.

As the benchmark criterion for the performance of an algorithm, we take its obtained accuracy of fit given a fixed amount of computational time.
Such a criterion corresponds to financial practice, since decisions based on derivative valuation calculations often need to be made within
seconds. To display the comparison results, we use the state-of-the-art and convenient \emph{performance profiles}; see \citeasnoun{dom02}. The
reader is referred there for details, but the idea is briefly described here. There are 100 test correlation matrices $p=1,\dots,100,$ and seven
algorithms $s=1,\dots,7$. As \emph{performance measure} we take the obtained function value $F^{(p,s)}$ of algorithm $s$ on problem $p$ given
the limited computational time. The \emph{performance ratio} $\rho^{(p,s)}$ is defined to be the ratio of the performance measure of the
algorithm over the best obtained performance measure for all seven algorithms,
\[
\rho^{(p,s)}=\frac{F^{(p,s)}}{\min\{F^{(p,s)}:s=1,\dots,7\}}.
\]
The cumulative distribution function $\Omega^{(s)}$ of the performance ratio for algorithm $s$, viewed as a random variable $p\to\rho^{(p,s)}$,
is then the \emph{performance profile} of algorithm $s$,
\[
\Omega^{(s)}(\xi) = \frac{1}{100} \# \{ p\; : \; \rho^{(p,s)}\le \xi,\; p=1,...,100 \}.
\]
A rule of thumb is, that the higher the profile of an algorithm, the better its performance. The performance profiles have been displayed in
Figures\begin{figure}[t]
\begin{center}
  \includegraphics[width=5in]{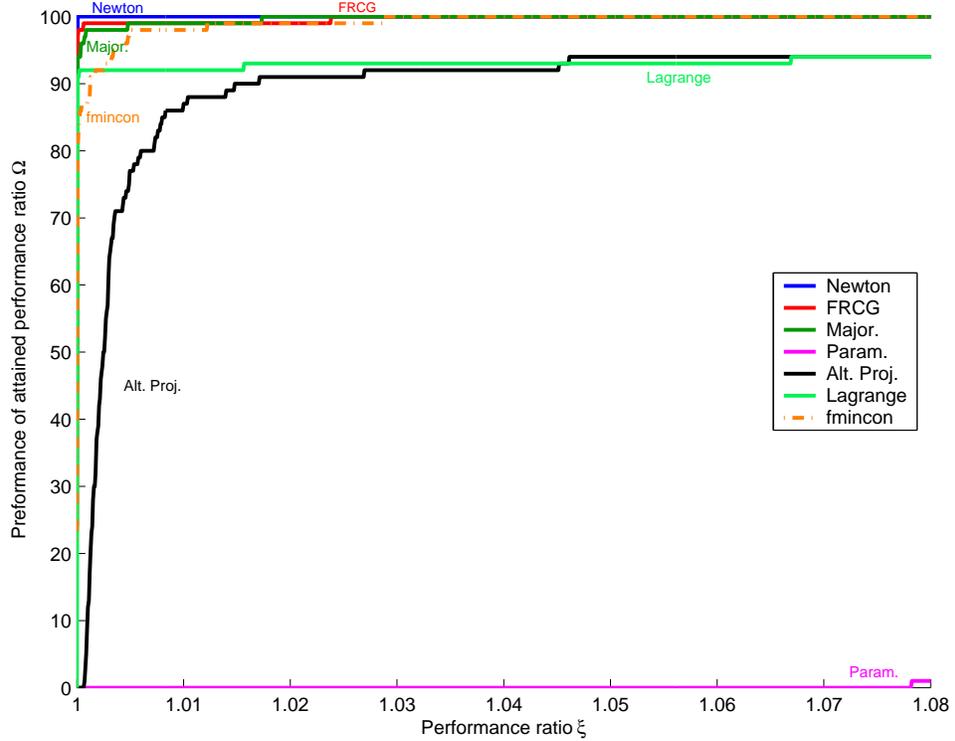}\\
  \caption{Performance profile with $n=30$, $d=3$, 2 seconds of computational time, Hadamard norm with equal weights. A rule of thumb is, that the higher the graph of an algorithm, the better its performance.}
   \label{fig:n30d3t2djdp}
  \end{center}
\end{figure}
\begin{figure}[t]
\begin{center}
  \includegraphics[width=5in]{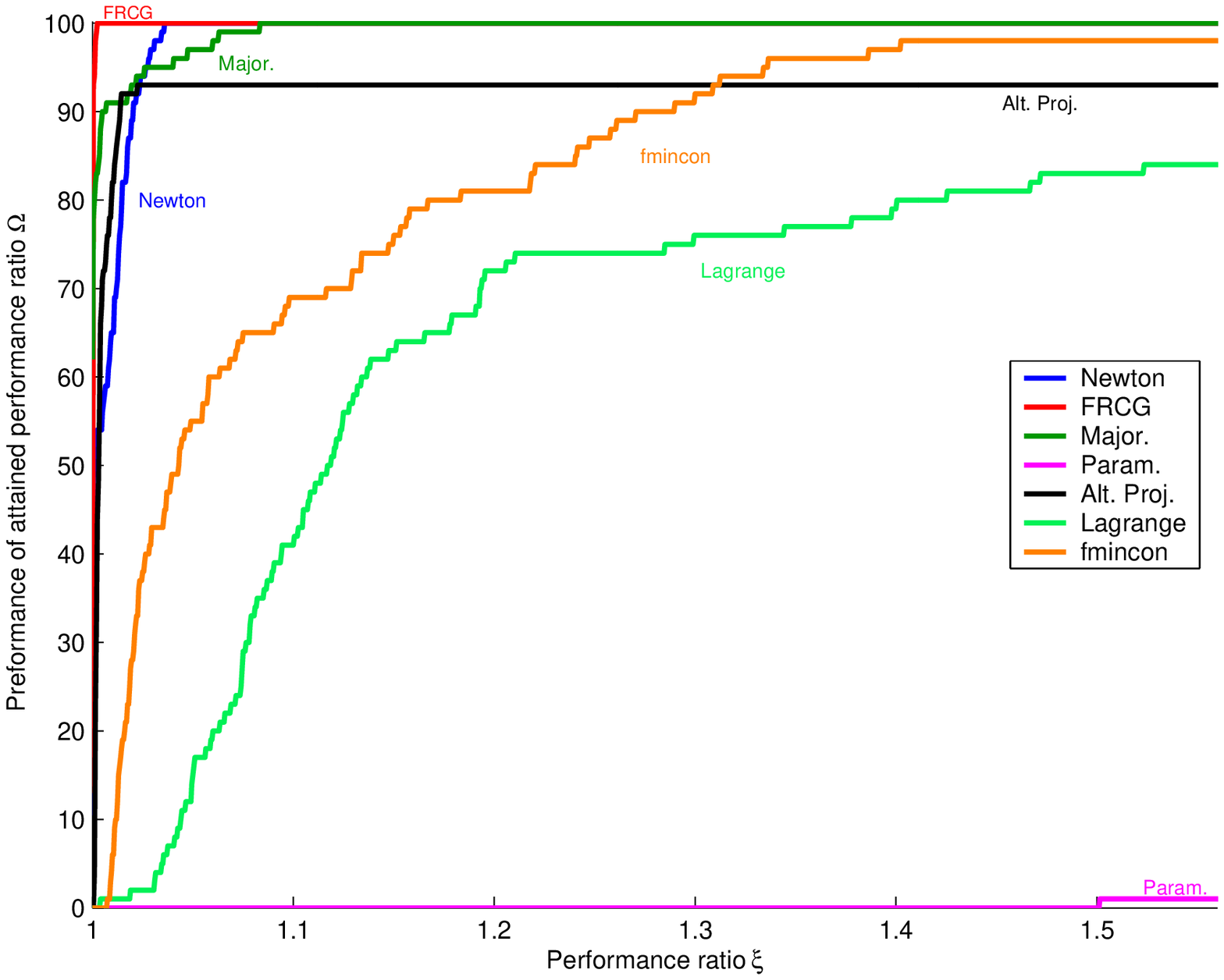}\\
  \caption{Performance profile with $n=50$, $d=4$, 1 second of computational time, Hadamard norm with equal weights.}
   \label{fig:n50d4t1djdp}
  \end{center}
\end{figure}
\begin{figure}[t]
\begin{center}
  \includegraphics[width=5in]{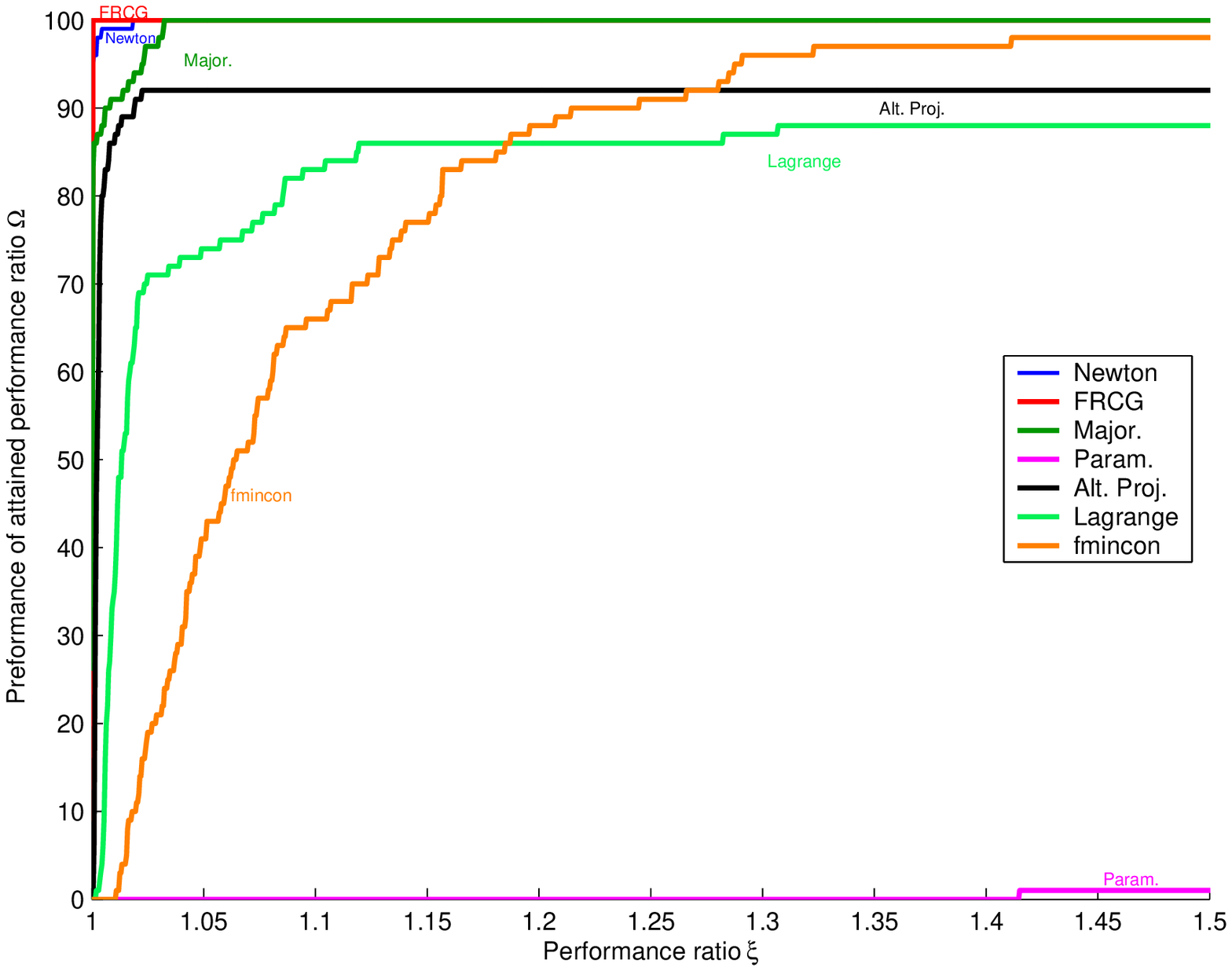}\\
  \caption{Performance profile with $n=60$, $d=5$, 3 seconds of computational time, Hadamard norm with equal weights.}
   \label{fig:n60d5t3djdp}
  \end{center}
\end{figure}
\begin{figure}[t]
\begin{center}
  \includegraphics[width=5in]{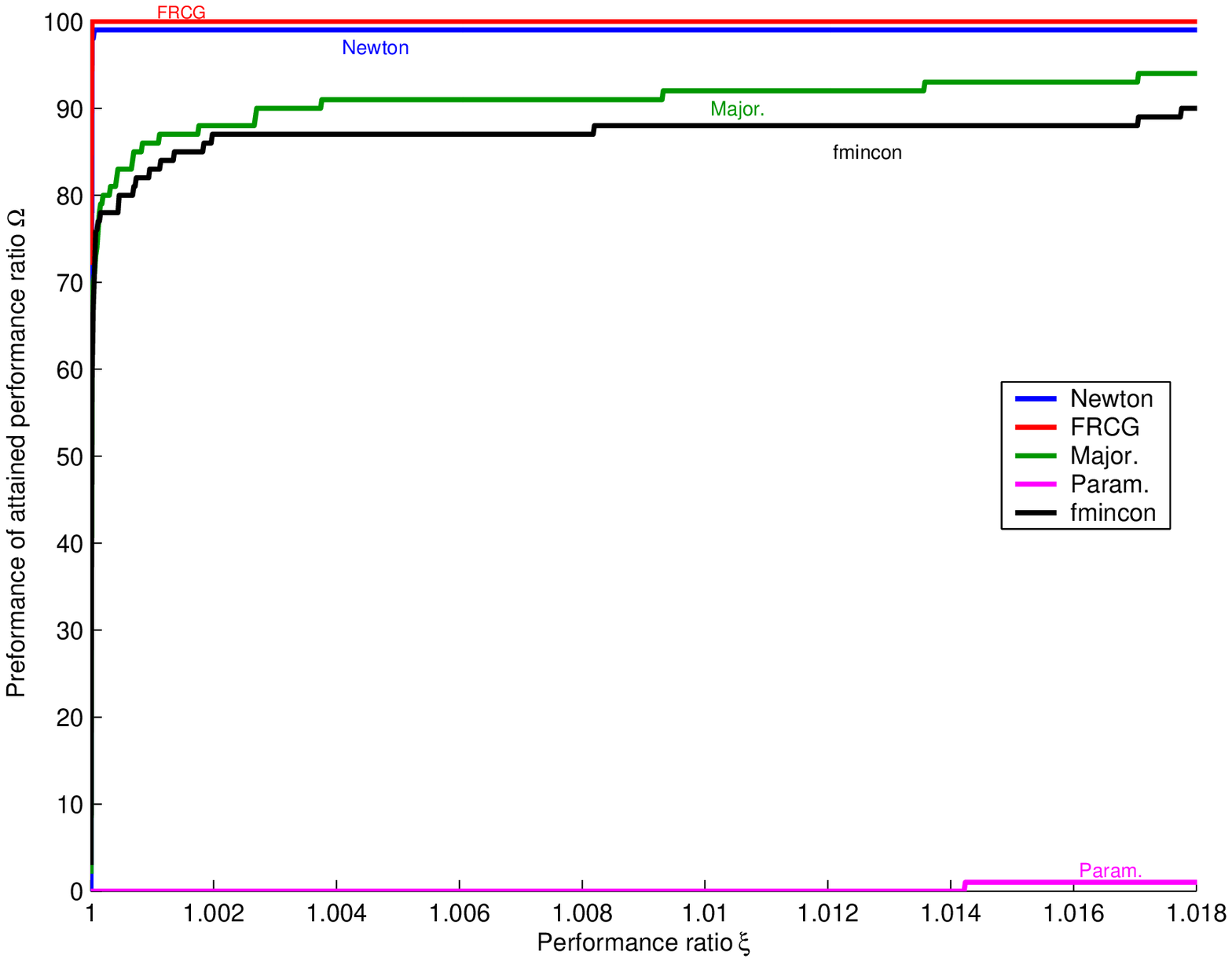}\\
  \caption{Performance profile with $n=15$, $d=3$, 1 second of computational time, trigger swap Hadamard semi-norm.}
   \label{fig:n15d3t1djdpWeightedHadamard}
  \end{center}
\end{figure}
\ref{fig:n30d3t2djdp}--\ref{fig:n15d3t1djdpWeightedHadamard}, for various choices of $n$, $d$, and computational times. Each performance profile
represents a benchmark on 100 different test interest rate correlation matrices. For Figures \ref{fig:n30d3t2djdp}--\ref{fig:n60d5t3djdp}, an
objective function with equal weights is used. For Figure \ref{fig:n15d3t1djdpWeightedHadamard}, we use a Hadamard semi-norm with non-constant
weights. These weights are chosen so as to reflect the importance of the correlation entries for a specific trigger swap, as outlined in, e.g.,
\citeasnoun[Section 20.4.3]{reb04fox}. For this specific trigger swap, the first three rows and columns are important. Therefore the weights
matrix $W$ takes the form
\[
W_{ij} = \left\{ \begin{array}{cl} 1 & \textrm{ if }i\le 3\textrm{ or }j\le 3, \\
0 & \textrm{ otherwise.} \end{array} \right.
\]

From Figures \ref{fig:n30d3t2djdp}--\ref{fig:n15d3t1djdpWeightedHadamard} it becomes clear that geometric optimisation compares favourably to
the other methods available in the literature, with respect to obtaining the best fit to the original correlation matrix within a limited
computational time.

\section{Conclusions} \label{sec:conclusions}

We applied geometric optimisation tools for finding the nearest low-rank correlation matrix. The differential geometric machinery provided us
with an algorithm more efficient than any existing algorithm in the literature, at least for the numerical cases considered. The geometric
approach also allows for insight and more intuition into the problem. We established a technique that allows one to straightforwardly identify
whether a local minimum is a global minimum.

\bibliographystyle{agsm} 
\bibliography{GeomRR}
\newpage
\appendix

\section{The proof of Theorem \ref{thm:the homeomorphism}} \label{sec:app proof thm the homeomorphism}

\emph{Proof of (i). The maps $\Psi$ and $\Phi$ are well defined: } To show that $\Psi$ is well defined, we need to show that if $Y_2\in[Y_1]$,
then $Y_2Y_2^T=Y_1Y_1^T$. From the assumption, we have that $\exists Q\in\Od:\;Y_2=Y_1 Q$. If follows that
\[
Y_2Y_2^T=(Y_1 Q)(Y_1 Q)^T=Y_1QQ^TY_1^T=Y_1Y_1^T,
\]
which was to be shown.

To show that $\Phi$ is well defined, we need to show:
\begin{enumerate}
\item[(A)] If $X\in\Cnd$ then there exists $Y\in\Tnd$ such that $X=YY^T$. \item[(B)] If $Y,Z\in\Tnd$, with $YY^T=ZZ^T=:X$ then there exists
$Q\in\Od$ such that $Y=ZQ$.
\end{enumerate}

Ad (A): Let
\[
X=Q\Lambda Q^T,\;\; Q\in O_n, \;\; \Lambda = \diag( \Lambda ),
\]
be an eigenvalue decomposition with $\Lambda_{ii}=0$ for $i=d+1,\dots,n$. Note that such a decomposition of the specified form is possible because of the
restriction $X\in\Cnd$. Then note that
\[
Q\sqrt{\Lambda} = \left( \begin{array}{c|c} (Q\sqrt{\Lambda})(:,1:d) & 0 \end{array} \right).
\]
Thus if we set $Y=(Q\sqrt{\Lambda})(:,1:d)$ then $YY^T=X$ and $Y\in\Tnd$, which was to be shown.

Ad (B): Let $\rank(Y)=\rank(Z)=\rank(X)=k\le d$. Without loss of generality, we may assume that the first $k$ rows of $Y$ and $Z$ are
independent. We extend the set of $k$ row vectors $\{Y_1,\dots,Y_k\}$ to a set of $d$ row vectors $\{Y_1,\dots,Y_k,\tY_{k+1},\dots,\tY_{d}\}$,
such that the latter forms a basis of $\real^{d}$. Similarly, we obtain a basis $\{Z_1,\dots,Z_k,\tZ_{k+1},\dots,\tZ_{d}\}$ of $\real^{d}$. It
follows that there exists an orthogonal rotation $Q$, $QQ^T=I_d$, such that $QY_i=Z_i$ $(i=1,\dots,k)$, $Q\tY_i=\tZ_i$ $(i=k+1,\dots,d)$. Note
that then also $QY_i=Z_i$ for $i=k+1,\dots,n$, by linearity of $Q$ and since the last $n-k$ row vectors are linearly dependent on the first $k$
row vectors by assumption. It follows that $YQ=Z$, which was to be shown.\hfill$\Box$

\emph{Proof (ii). Diagram (\ref{eq:the diagram}) is commutative: } To show that $\Psi\circ\pi=s$: Let $Y\in\Tnd$, then $\pi(Y)=[Y]$ and
$\Psi([Y])=YY^T$ and also $s(Y)=YY^T$. To show that $\Phi\circ s=\pi$: Let $Y\in\Tnd$, then $s(Y)=YY^T$ and $\Phi(YY^T)=[Y]$ and also
$\pi(Y)=[Y]$.\hfill$\Box$

\emph{Proof of (iii). The map $\Psi$ is a homeomorphism with inverse $\Phi$: } It is straightforward to verify that $\Phi\circ\Psi$ and
$\Psi\circ\Phi$ are both the identity maps. The map $\Psi$ is thus bijective with inverse $\Phi$. To show that $\Psi$ is continuous, note that
for quotient spaces we have: The map $\Psi$ is continuous if and only if $\Psi\circ\pi$ is continuous (see for example \citeasnoun{amr88},
Proposition 1.4.8). In our case, $\Psi\circ\pi=s$ with $s(Y)=YY^T$ is continuous. The proof now follows from a well-known lemma from topology: A
continuous bijection from a compact space into a Hausdorff space is a homeomorphism (see for example \citeasnoun{mun75}, Theorem
5.6).\hfill$\Box$

\section{Proof of Proposition \ref{prop:free part is manifold}} \label{sec:app proof free part manifold}

\begin{enumerate}

\item[1.] It is sufficient to show that $\{Y\in\real^{n \by d}:\rank(Y)=d\}$ is open in $\real^{n \by d}$, since $\Tnd^*$ is open in $\Tnd$ if
and only if $\{Y\in\real^{n \by d}:\rank(Y)=d\}$ is open in $\real^{n \by d}$. Since the rank of a symmetric matrix is a locally constant
function, it follows that $\{Y\in\real^{n \by d}:\rank(Y)=d\}=s^{-1}(\rank^{-1}(d))$ is an open subset of $\real^{n \by d}$, with $s(Y)=YY^T$ as
in Definition \ref{def:Snd over Od and the maps}. \hfill$\Box$

\item[2.] This part is a corollary of Theorem 1.11.4 of \citeasnoun{duk00}. This theorem essentially states that for a smooth action of a Lie
group on a manifold the quotient is a manifold if the action is proper and free. First, we show that the action of $O_d$ on $\Tnd^*$ is
proper\footnote{For a definition, see \citeasnoun[page 53]{duk00}.}.  Let
\[
\phi :  \Tnd^* \times O_d  \rightarrow   \Tnd^* \times \Tnd^*, \;\; (Y,Q) \mapsto  (YQ,Y)
\]
and $K$ a compact subset of $\Tnd^* \times \Tnd^*$. We have to show that $\phi^{-1}(K)$ is compact.  By continuity of $\phi$, $\phi^{-1}(K)$ is
closed in $\Tnd^* \times O_d$. Because $\Tnd^* \times O_d$ is bounded it follows that $\phi^{-1}(K)$ is compact.

Second, we show that the $O_d$-action on $\Tnd^*$ is free. Let $Y \in \Tnd^*$ and $Q \in O_d$ such that $YQ=Y$. Since $\rank(Y)=d$, it follows
from the proof of Theorem \ref{thm:the homeomorphism} ($i$) that there exists precisely one $Q \in \Od$ such that $YQ=Y$. Thus, this $Q$ must be
the identity matrix.

The dimension of $M_{n,d}^*=\dim(\Tnd^*)-\dim(O_d)=n(d-1)-\half d(d-1)$. \hfill$\Box$
\end{enumerate}

\section{Proof of Proposition \ref{prop:diffstructureMnd}} \label{sec:app proof diffstructureMnd}

This part is a corollary of Theorem 1.11.4 of \citeasnoun{duk00}. This theorem states that there is only one differentiable structure on the
orbit space which satisfies the following: Suppose that, for every $[Y] \in \Mnd^*$, we have an open neighbourhood $\calu \subseteq \Mnd^*$ and
a bijective map:
\[
\tau:\pi^{-1}(\calu) \to \calu \times \Od, \;\; Y \mapsto (\pi(Y),\chi(Y)),
\]
such that, for every $Y \in \pi^{-1}(\calu) $, $Q \in  \Od$, $\tau( YQ)=(\pi(Y),\chi(Y)Q)$. The differentiable structure on $\R$ is the one
which makes $\tau$ into a diffeomorphism. The topology of $\Mnd^*$ obtained in this manner is equal to the quotient topology.

Let $Y \in \Pu$ and  $\sigma_Y$  be a section over $\calu_Y$ defined in (\ref{eq:section}). We define $\tau_Y:\pi^{-1}(\mathcal{U}_Y)\to
\mathcal{U}_Y \by \Od $ as follows. For $Z \in \pi^{-1}([Z])$, $[Z] \in \mathcal{U}_Y$, there is a unique element $Q_Z \in \Od$ such that
$Z=\sigma_Y([Z])Q_Z$. Then we define $\tau_Y$ by $\tau_Y(Z)=([Z],Q_Z)$. By definition, we have that $\tau^{-1}_Y([Z],Q)=\sigma_{Y}([Z])Q$. Since
$\tau^{-1}_Y:\pi^{-1}(\mathcal{U}_Y) \to \mathcal{U}_Y \by \Od$ is a bijective map, we have that $\tau_Y$ is bijective, too. It can be easily
verified that $\tau_Y$ satisfies the condition $\tau_Y(YQ)=([Y],Q_YQ)$ of Theorem 1.11.4 of \citeasnoun{duk00} stated above. It follows that
$\tau_Y$ is a diffeomorphism if and only if $\sigma_Y:\calu\to\sigma_Y(\calu)$ is a diffeomorphism. Thus, the differentiable structure on $\R$
is the one which makes $\sigma_Y:\mathcal{U}_Y \to \sigma{(\mathcal{U}_Y)}$ into a diffeomorphism.\hfill$\Box$

\section{Proof of Theorem \ref{thm:surjective}} \label{sec:app proof surjective}

Let $X\in\Cnd$ and suppose that $\rank(X)=k\le d$. Then there is a $Y\in T_{n,k}$ such that $YY^T=X$, by Theorem
\ref{thm:the homeomorphism}. Apply to $Y$ the procedure\footnote{The procedure in Section
\ref{sec:differentiable structure} is stated in terms of $d$, but $k$ should be read there in this case.}
outlined in Section \ref{sec:differentiable structure}, to obtain a lower-triangular matrix $Y^*\in T_{n,k}$,
such that $Y^*(Y^*)^T=X$. A lower-triangular matrix $\Yb\in\cholnd$ that satisfies $\Yb\Yb^T=X$ can now easily
be obtained by setting
\[
\Yb=\left( \begin{array}{cc} Y^{*} & \underbrace{0}_{n\times(d-k)} \end{array} \right),
\]
which was to be shown.\hfill$\Box$

\section{Proof of Theorem \ref{thm:loc min on chol is loc min on corr}} \label{sec:app proof thm loc min on chol is loc min on corr}

First, we prove the `only if' part. Note that it is sufficient to show that the map $s:\cholnd \to\Cnd$ is \emph{open}. For then if $Y$ attains
a local minimum of $F$ on the open neighbourhood $\calu\subset\cholnd$, then $s(Y)=YY^T$ attains a local minimum of $\tilde{F}$ on the open
neighbourhood $s(\calu)$ of $YY^T$, since for any $X'=Y'Y'^T\in s(\calu)$, $\tilde{F}(X')=\tilde{F}(Y'Y'^T)=F(Y')\ge F(Y)=\tilde{F}(YY^T)$.

To show that $s:\cholnd\to\Cnd$ is open, note that it is sufficient to show that $\pi:\cholnd\to\Mnd$ is open, since $\Psi:\Mnd\to\Cnd$ is a
homeomorphism (see Proposition \ref{prop:free part is manifold}, item 3) and $s=\Psi\circ\pi$.

Suppose, then, that $\calu$ is open in $\cholnd$. We have to show that $\pi^{-1}( \pi(\calu))$ is open in $\Tnd$, by definition of the quotient
topology of $\Mnd$. We have
\[
\pi^{-1}\big( \ \pi(\calu) \ \big) = \big\{ \ YQ \ : \ Y\in\calu, \ Q\in O_d \ \big\} .
\]
It is sufficient to show that the complement $(\pi^{-1}(  \pi(\calu) ))^{c}$ is closed. Let $\{Y^{(i)}\}$ be a sequence in $(\pi^{-1}(
\pi(\calu) ))^{c}$ converging to $Y$, i.e. $\lim_{i \to \infty} \|Y^{(i)} - Y \| = 0$. We can write $Y^{(i)}= Z^{(i)} Q^{(i)}$ with $Z^{(i)} \in
\calu^{c}$ and $Q^{(i)} \in \Od$. Then,
\begin{equation} \label{eq:lim Yi-Y}
\lim_{i \to \infty} \|Y^{(i)} - Y \|= \lim_{i \to \infty}\|Z^{(i)} Q^{(i)} - Y \| = \lim_{i \to \infty} \|Z^{(i)}  - Y (Q^{(i)})^T \|=0.
\end{equation}
Since $\calu^c\times O_d$ is compact, there exists a convergent subsequence $\{(Z^{(i_j)},Q^{(i_j)})\}$, with $Z^{(i_j)}\to Z^*\in \calu^{c}$
and $Q^{(i_j)}\to Q^*$, say. From (\ref{eq:lim Yi-Y}) it follows that $Z^*=Y (Q^*)^T  \in \calu^{c}$, which implies $Y \in (\pi^{-1}( \pi(\calu)
))^{c}$.

The reverse direction is obvious since the map $s:\cholnd\to\Cnd$ is continuous.\hfill$\Box$

\section{Proof of Lemma \ref{lem:connection normal vs lagrange}} \label{sec:app proof connection normal lagrange}

It is recalled from matrix analysis that $X_1$ and $X_2$ admit a joint eigenvalue decomposition if and only if their Lie bracket
$[X_1,X_2]=X_1X_2-X_2X_1$ equals zero. Define $\bar{C}(\lambda):=-\psi+\lambda$. Note that $2 \lambda Y$ is the projection $\pi_{N_Y\Tnd}(F_Y)$
of $F_Y$ onto the normal space at $Y$. Note also that
\begin{equation} \label{eq:feasibility}
YY^T+\bar{C}(\lambda)=C(\lambda).
\end{equation}
We calculate
\begin{equation} \label{eq:complementary slackness}
\bar{C}(\lambda) Y = \big\{  -\psi+\lambda \; \big\}Y = -\half F_Y + \half \pi_{N_Y\Tnd}\big( \; F_Y \; \big) = 0.
\end{equation}
The last equality follows from the assumption that  $\Grad F(Y)=0$, i.e.~the differential $F_Y$ is normal at $Y$. (Here, $\Grad F(Y)$ denotes
the gradient on $\Tnd$.) It follows from (\ref{eq:complementary slackness}) and from the symmetry of $\bar{C}(\lambda)$ that
\begin{enumerate}
\item[(i)] $YY^T \bar{C}(\lambda) = 0$ and also, \item[(ii)] $[YY^T,\bar{C}(\lambda)]=0$.
\end{enumerate}
From (ii), $YY^T$ and $\bar{C}(\lambda)$ admit a joint eigenvalue decomposition, but then also jointly with $C(\lambda)$ because of
(\ref{eq:feasibility}). Suppose $\bar{C}(\lambda)=Q\bar{D}Q^T$. From (i) we then have that $D_{ii}^*$ and $\bar{D}_{ii}$ cannot both be
non-zero. The result now follows since $YY^T$ is positive semidefinite and has rank less than or equal to $d$. \hfill$\Box$

\section{Proof of Theorem \ref{thm:global minimum}} \label{sec:app proof thm global minimum}

Define the Lagrangian
\[
\call(X,\lambda) := -\| C-X \|_F^2 -2\lambda^T\diag(C-X), \;\; \textrm{ and}
\]
\begin{equation} \label{eq:V min}
V(\lambda) := \min\big\{ \; \call(X,\lambda) \; : \; \rank(X) = d \; \big\}.
\end{equation}
Note that the minimization problem in Equation (\ref{eq:V min}) is attained by any $\{C(\lambda)\}_d$ (see e.g.,
Equation (30) of \citeasnoun{wu03}). For any $X\in\Cnd$,
\[
\|C-X\|_F^2 \stackrel{(a)}{=} -\call(X,\lambda^*) \stackrel{(b)}{\ge} -V(\lambda^*) \stackrel{(c)}{=} \|
C-\{C(\lambda)\}_d \|_F^2.
\]
(This is the equation at the end of the proof of Theorem 4.4 of \citeasnoun{zhw03}.) Here (in-)equality
\begin{enumerate}
\item[(a)] is obtained from the property that $X\in\Cnd$, \item[(b)] is by definition of $V$, and \item[(c)] is by assumption of (\ref{eq:assume
diag C+l_d = diag C}).\hfill$\Box$
\end{enumerate}

\phantom{\citeasnoun{rebb99}\citeasnoun{reb02}\citeasnoun{reb99jcf}\citeasnoun{reb04fox}}

\end{document}